\documentclass{optica-article}

\journal{opticajournal} 

\articletype{Research Article}

\begin{document}

\title{Laser‑Stabilised Ionising Transitions}

\author{Erika Cortese,\authormark{1,*}  and Simone De Liberato\authormark{1,2}}

\address{\authormark{1}School of Physics and Astronomy, University of Southampton, Southampton SO17 1BJ, United Kingdom\\

\authormark{2}Istituto di Fotonica e Nanotecnologie -- Consiglio Nazionale delle Ricerche (CNR), Piazza Leonardo da Vinci 32, Milano, Italy}

\email{\authormark{*}E.Cortese@soton.ac.uk} 


\begin{abstract*} 
We investigate a ionising electronic transition under resonant pumping. We demonstrate that, above a critical value of the pump intensity, a novel metastable electronic bound state is created, which can decay into the free electron continuum by two-photon ionization. We calculate the system's resonant fluorescence spectrum, finding results qualitatively different from the Mollow triplet expected in a bound-to-bound transition. The fluorescent emission can be used to measure the time-resolved population of the novel metastable state.
Contrary to Kramers-Hennenberger atoms, stabilised by non-perturbative, non-resonant laser pulses, the physics we observe is inherently resonant and relies on perturbative level repulsion. In analogy to how the AC-Stark shift is a semiclassical version of the single-photon Rabi splitting observed in photonic cavity, the phenomenon we describe is better understood as a semiclassical version of recently observed excitons bound by a single cavity photon.
\end{abstract*}

\section{Introduction}
 \label{sec:1}
Understanding the interaction between bound states and continuum free-particle states is crucial to describe a wide range of physical phenomena, including ionisation, dissociation, and energy loss. In atomic and molecular systems near the ionisation threshold, this interplay governs photoexcitation dynamics, giving rise to the Fano asymmetric line shape in photoionisation spectra, an effect resulting from quantum interference between bound and continuum states coupled via direct laser excitation \cite{fano_effects_1961,litvinenko_multi-photon_2021}.
Beyond atomic physics, the Fano theory has been extended to 
bulk \cite{huttner_quantization_1992} and cavity-embedded \cite{de_liberato_virtual_2017} dielectrics, 
where bound-to-continuum transitions are central to mid-infrared optoelectronic technologies as quantum well infrared photodetectors (QWIPs) \cite{sirtori_nonparabolicity_1994,vigneron_quantum_2019} and quantum cascade lasers \cite{faist_quantum-cascade_2001,hofstetter_quantum-cascade-laser_2002,pflugl_high-temperature_2003}.
\begin{figure}
    \centering
\includegraphics[width=0.8\columnwidth]{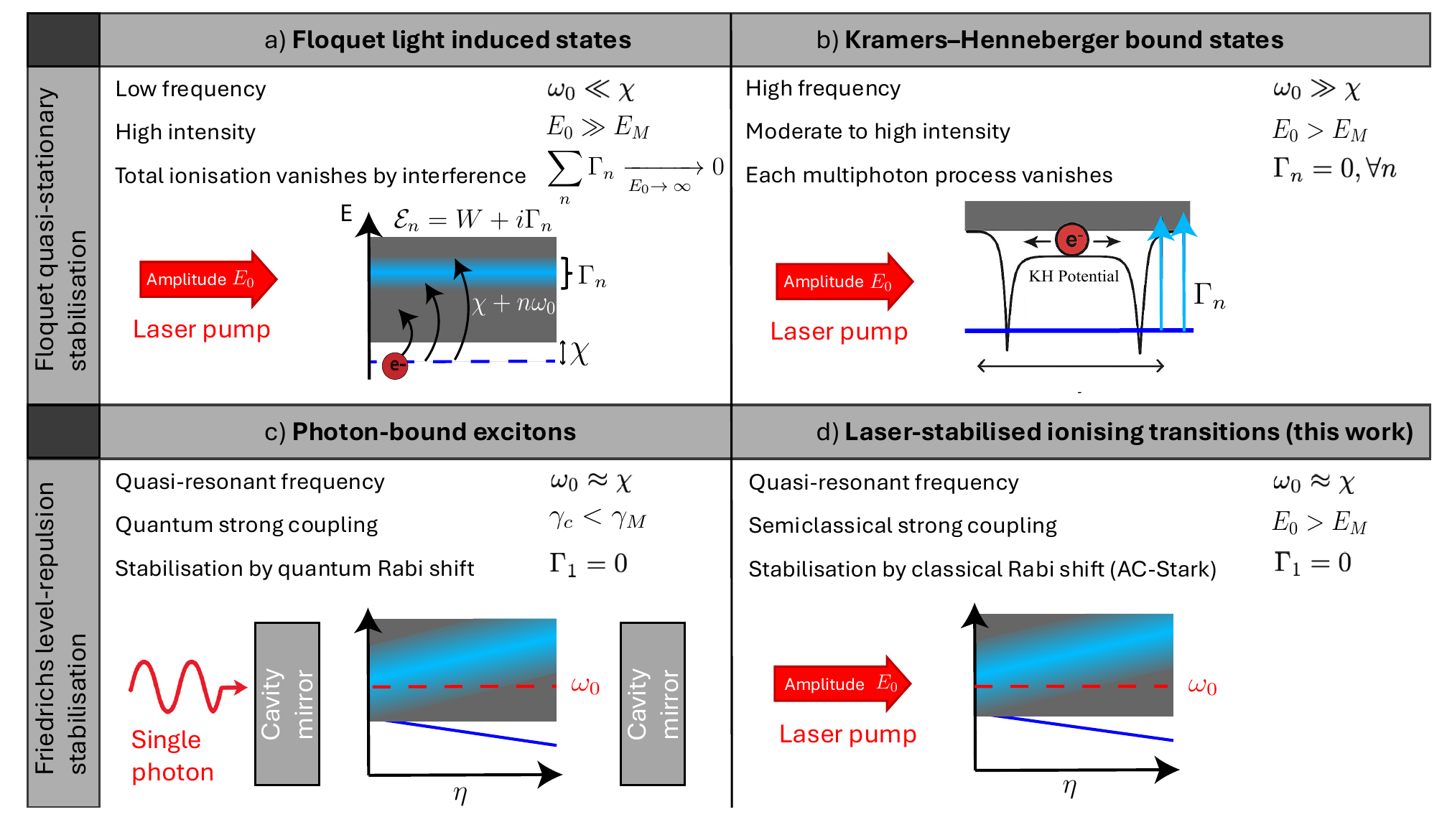}
    \caption{\label{Fig:0} \textbf{Different Light-stabilised Frameworks.}
(a) Schematic representation of Floquet quasistationary (QS) stabilisation for a single bound state of energy 
$-\chi$ (blue dashed line), nonperturbatively photoexcited by a coherent drive of frequency $\omega_0$ into the continuum 
via multiphoton channel interference in the low-frequency, high-intensity regime (laser amplitude $E_0$ much larger than a critical amplitude $E_M$ depending on the parameters of the matter system). 
The resulting Floquet–Gamow quasienergies $\mathcal{E}_n$ possess a real part $W$ and an imaginary part $\Gamma_n$, corresponding 
to the $n$-photon ionization rate (indicated by the width of the cyan shaded regions). Destructive interference among 
different photon channels leads to a total ionization rate $\Gamma = \sum_n \Gamma_n$ that decays to zero in an oscillatory manner at increasing amplitude. 
(b) High-frequency limit illustrated in the Kramers--Henneberger (KH) frame. For moderate to large $E_0$, the effective potential develops new minima that support light-induced bound states 
(blue solid lines). Expanding the laser-dressed potential, only the zeroth-order term, referred to as KH potential,
 dominates, while multiphoton processes mediated by higher-order terms 
become negligible 
($\Gamma_n \rightarrow 0$). 
(c) Photon-bound exciton (PBE, blue solid line) emerging from resonant level repulsion between the ionising transition and a photonic resonator as light-matter normalised coupling $\eta$ increases.
The discrete resonance, protected by single-photon ionization, becomes visible in the strong coupling regime when the cavity linewidth $\gamma_c$ is smaller than a critical linewidth $\gamma_M$ depending on the parameters of the matter system. 
(d) Semiclassical analogue of the PBE, under investigation in this work, where resonant level repulsion occurs between coherently driven ionizing transitions as the parameter quantifying the coupling strength $\eta$ increases.}
\end{figure}

\begin{figure}
    \centering
\includegraphics[width=0.6\columnwidth]{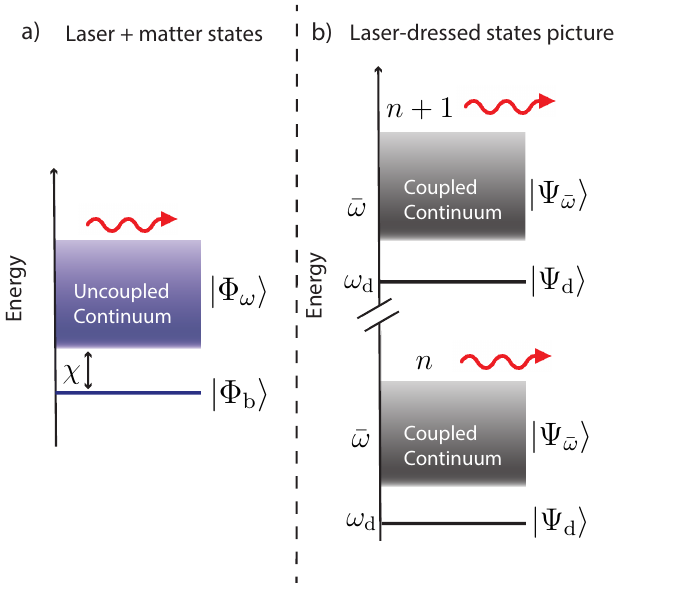}
    \caption{\label{Fig:1} \textbf{Light-dressed states framework.}  (a) Schematic representation of a single bound state $|\Phi_{\text{b}}\rangle$  coupled to a continuum of free states $|\Phi_\omega\rangle$  (blue shaded region)  via a semiclassical monochromatic laser field, with the ionisation threshold $\chi$. (b) Corresponding light-dressed states, where the discrete state $|\Psi_{\text{d}}\rangle$ and the continuum of states $|\Psi_{\bar{\omega}}\rangle$  (black shaded region) are arranged in adjacent photon ladders, characterised by photon numbers $n$ and $n+1$. }
\end{figure}

Although not often emphatised, the Fano approach to diagonalise bound-to-continuum transitions relies on the crucial hypothesis that the spectrum of the coupled system is entirely continuum. A more general mathematical treatment shows that, due to the non-analyticity of the spectrum of the Lee–Friedrichs Hamiltonian, discrete bound states can appear outside of the continuum for large enough coupling strengths \cite{passante_bound_2000}. In solid-state cavity quantum electrodynamics devices, these states can be interpreted as the result of the off-resonant strong coupling between the discrete and continuum parts of the spectrum \cite{cortese_strong_2019,cortese_exact_2022}, providing a photonic-based approach to modify the properties of material embedded in photonic resonators \cite{khurgin_excitonic_2001,brodbeck_experimental_2017}. 
These states have been recently observed in doped quantum wells \cite{cortese_excitons_2021,rajabali_polaritonic_2021}, where they correspond to photon-bound excitons (PBEs), metastable bound states of the electron gas stabilised by their entanglement with a cavity photon \cite{kumar_microscopic_2022}. They are metastable because of the finite lifetime of the cavity photon binding the electrons.

The concept of stabilisation of atoms exposed to intense laser fields has been an important topic in atomic physics for decades, revealing that, under certain conditions, increasing the field intensity can counterintuitively reduce ionisation. Over the decades, several theoretical frameworks have been developed to describe this effect, each valid in a particular parameter regime, as illustrated in Fig.~\ref{Fig:0}. 

In the Floquet quasistationary (QS) picture (top panels of Fig.~\ref{Fig:0}), applicable both in the low-frequency--high-intensity (panel a) and high-frequency--moderate-intensity (panel b), the bound-to-continuum coupling generally occurs via multiphoton ionisation channels. However, the underlying stabilisation mechanism, often referred to \textit{adiabatic stabilisation},  arises from different physical processes depending on the parameter regime. In the low-frequency/high-intensity limit, 
stabilisation arises when destructive interference among multiphoton ionisation pathways suppresses the net ionisation rate, giving rise to nearly stationary field-dressed states \cite{gavrila2002atomic,blekher1992floquet,mori2023floquet}. These light-induced stationary states can thus be regarded as dynamical analogues of bound states in the continuum (BICs) in photonic and electronic systems, where, instead of spatial destructive interference or symmetry, it is temporal interference that suppresses coupling to free-state channels despite energetic overlap with the continuum~\cite{hsu2016bound}.

When instead the laser frequency exceeds the binding energy, the electron experiences an effectively time-averaged potential, and the mechanism acquires a more transparent interpretation. In the high-frequency limit, a complementary and spatially intuitive interpretation is provided by the Kramers--Henneberger (KH) frame \cite{popov1999applicability, pawlak2014conditions, henneberger_perturbation_1968, matthews_amplification_2018}, where the oscillating electric field is transformed away and the electron experiences a quiver-averaged potential, where quiver amplitude refers to the maximum displacement of the electron during its oscillatory motion in the laser field (Fig. \ref{Fig:0} b). At sufficiently large field amplitude, this potential develops additional minima capable of supporting light-induced bound states. These KH states offer a real-space picture of the same stabilisation phenomenon described by Floquet theory: the electron becomes localized in a field-dressed potential that couples only weakly to the continuum. In this limit, the process remains essentially adiabatic, since the electron dynamics follow the slowly varying envelope of the effective potential.

KH light-induced bound states and PBEs (schematically represented in Fig. \ref{Fig:0} c)  thus emerge in fundamentally different regimes of light-matter coupling: the former in a semiclassical, off-resonant, nonperturbative coherent field, the latter in a \textit{quantum} regime characterized by quasi-resonant coupling to a single photon. Likewise, the physical mechanisms underpinning stabilisation are distinct: for KH states, it results from the cycle-averaged modification of the binding potential in a rapidly oscillating field, for a PBE, it arises from quasi-resonant strong coupling, where level repulsion between a discrete resonance and the continuum pushes the hybridized state below the ionisation threshold. 

Despite these differences, it is natural to ask whether a new form of laser-stabilised bound state could emerge in an intermediate regime, that is by resonantly driving an ionizing transition (\ref{Fig:0} d), but through a mechanism analogous to that producing PBE. The mechanism we propose is related to PBE in the same way semiclassical AC-Stark splitting and Rabi oscillations in the semiclassical Rabi model are related to vacuum Rabi oscillations in the the quantum Rabi model~\cite{de_bernardis_tutorial_2024}. 

Resonant fluorescence is a standard probe of the coherent dynamics of quantum emitters, with even the simplest two-level system exhibiting the rich structure of the Mollow triplet~\cite{mollow_power_1969}, observed across a broad range of atomic, molecular, and solid-state platforms~\cite{ortiz-gutierrez_mollow_2019,ng_observation_2022,shammah_terahertz_2014,ulhaq_detuning-dependent_2013,moelbjerg_resonance_2012}. In the present work, we show that this framework can also reveal the existence of laser-stabilised resonances, which appear as narrow features in the resonance fluorescence spectrum of a resonantly pumped ionizing transition. In our model, these resonances are metastable: spontaneous emission processes that produce the resonance fluorescence also provide irreversible decay channels that eventually deplete the stabilised population into the continuum of ionized states.

For clarity, the detailed derivations and analytical steps are collected in the Appendices, while the main text presents only the essential results and physical reasoning needed to follow our arguments.

\begin{figure*}
    \centering \includegraphics[width=\linewidth]{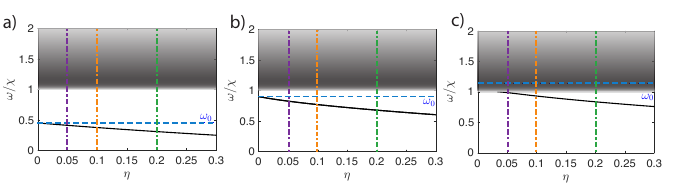}
    \caption{\label{Fig:2} \textbf{Quasi-classical laser-stabilised discrete resonances.}
Dispersion of the dressed modes as a function of the normalised coupling strength for three pumping regimes: below two-photon threshold ($\omega_{0}<\chi/2$, a), below threshold ($\chi/2<\omega_{0}<\chi$, b), and above threshold ($\omega_{0}>\chi$, c). The solid black line tracks the discrete level $\omega_{\text{d}}$; the black-shaded region marks the continuum that begins at the edge of the continuum $\chi$, with the internal grey scale indicating how the dipole moment varies across the continuum. The blue dashed line shows the laser-pump frequency $\omega_{0}$. The purple, orange, and green vertical dash-dotted lines mark three values of the coupling which will be used as reference in the rest of the paper.}
\end{figure*}
\begin{figure*}
    \centering \includegraphics[width=\linewidth]{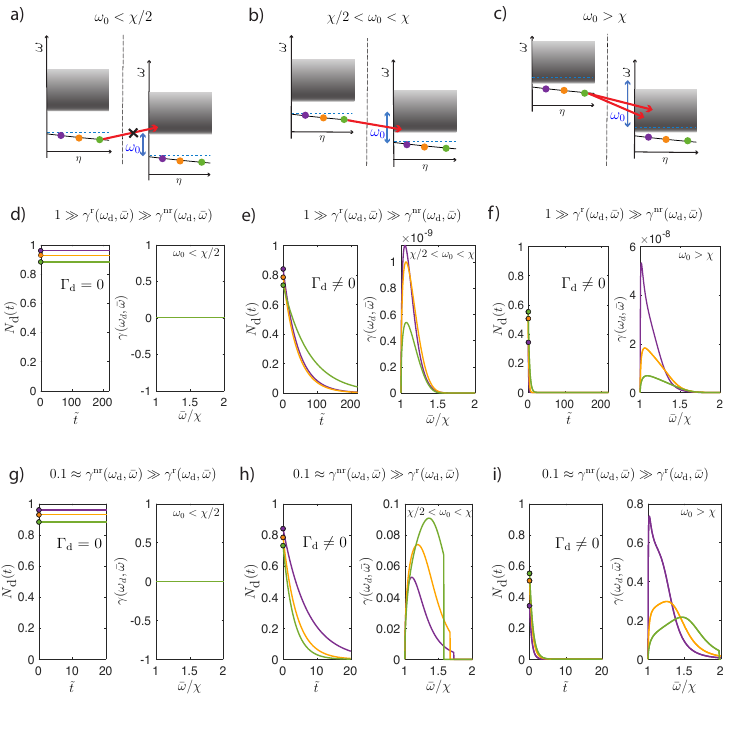}
    \caption{\label{Fig:3} \textbf{Dynamics of the discrete-state population.}
The panels in the top row (a-c) schematically sketch two replica of the spectrum of the system separated by the energy of a laser photon. In all the present Figure purple, orange, and green colors mark the values of the normalised coupling highlighted in Fig.~\ref{Fig:2}. The red arrows indicate the allowed decay channels in the case $\eta=0.2$ (green).
Each of the other panels (d-i) represents on the left the time evolution of the discrete-state population $N_{\text{d}}(t)$ and on the right the total discrete-to-continuum emission rate $\gamma(\omega_{\text{d}},\bar{\omega})$. Each column corresponds to a different value of the pump frequency, corresponding to the three cases shown in Fig.~\ref{Fig:2}: $\omega_{0}<\chi/2$ (d,g), $\chi/2<\omega_{0}<\chi$ (e,h), and $\omega_{0}>\chi$ (f,i). 
The second row (d-f) describes a system where a weak radiative loss channel dominates $(1 \gg \gamma^{\text{r}}(\omega_{\text{d}},\bar{\omega})\gg\gamma^{\text{nr}}(\omega_{\text{d}},\bar{\omega})) $, while the third (g-i) a system where a large  non-radiative channel dominates $(0.1\approx\gamma^{\text{nr}}(\omega_{\text{d}},\bar{\omega})\gg\gamma^{\text{r}}(\omega_{\text{d}},\bar{\omega}) )$. All the times have been normalised over the maximum value of the decay rate over the three values of the normalised couplings and the laser frequencies considered $\tilde{t}=t \max_{\eta=\{0.05,0.1,0.2\}, \omega_0} \Gamma_{\text{d}}$.}
\end{figure*}

\begin{figure*}
    \centering
\includegraphics[width=0.95\textwidth]{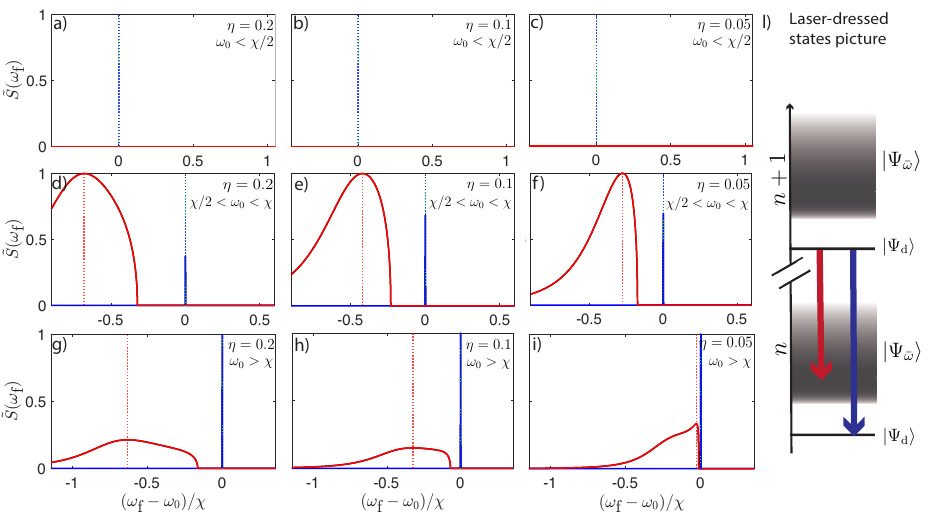}
    \caption{\label{Fig:4}\textbf{ Dipole fluorescence spectra for the for dominant radiative losses} ($1\gg\gamma^{\text{r}}(\omega_{\text{d}}, \bar{\omega})\gg\gamma^{\text{nr}}(\omega_{\text{d}}, \bar{\omega}) $):
Panels (a–c) show the calculated spectra $\tilde{S}(\omega_\textrm{f}) $ for the below two-photon threshold regime
$(\omega_{0} < \chi/2)$; panels (d–f) for the below-threshold regime
$(\chi/2 < \omega_{0} < \chi)$; and panels (g–i) for the above-threshold
regime $(\omega_{0} > \chi)$.
The tilde   $\tilde{}$  denotes the normalisation of $S(\omega_\textrm{f})$  to its peak amplitude in each panel (including any unphysical negative frequency peaks). 
Each column corresponds to a different normalised coupling strength
$\eta$: $\eta = 0.20$ in panels (a, d, g), $\eta = 0.10$ in
(b, e, h), and $\eta = 0.05$ in (c, f, i).
In every spectrum the red curve represents the discrete-to-continuum
component $S_{\text{d-c}}(\omega_{\text{f}})$, while the blue curve shows the discrete-to-discrete component $S_{\text{d-d}}(\omega_{\text{f}})$.
The blue dashed vertical line marks the laser energy $\omega_{0}$, and
the red dashed vertical line indicates the energy at which the
hybridisation  $|c_{\text{b}}(\bar{\omega})|^2$ reaches its maximum.
Curve colors follow the same coding used in panel (l), which sketches
the decay channels in the dressed-state picture.
   }
\end{figure*}

\begin{figure*}
    \centering
\includegraphics[width=0.95\textwidth]{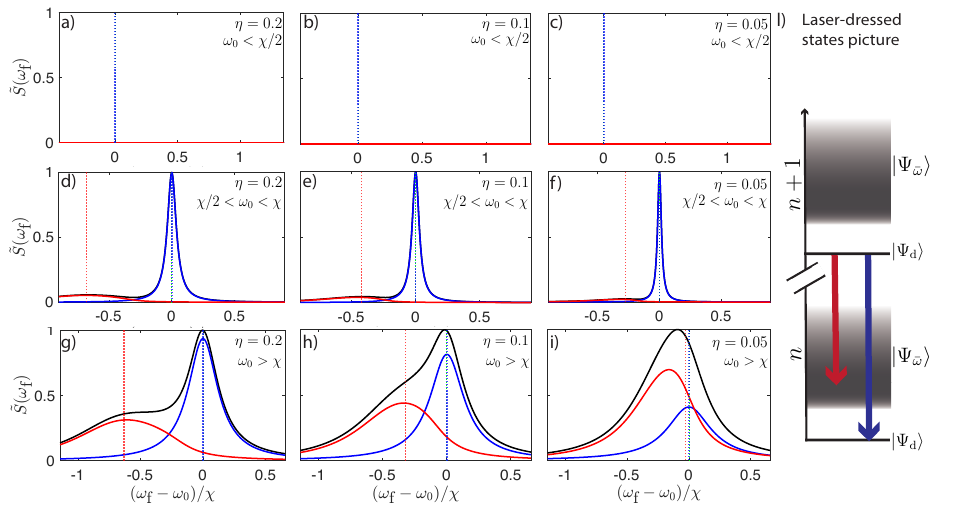}
    \caption{\label{Fig:5}\textbf{ Dipole fluorescence spectra for dominant non-radiative losses }
    ($0.1\approx\gamma^{\text{nr}}(\omega_{\text{d}}, \bar{\omega})\gg\gamma^{\text{r}}(\omega_{\text{d}}, \bar{\omega})$).
    Same as in Fig.~\ref{Fig:4} but in the case of dominant non-radiative losses, non-negligible comparing to the binding energy. Details on the numerical parameters can be found in the Appendix \ref{appD}. }
\end{figure*} 
\section{Theoretical model}\label{sec:2}
\subsection{Problem setup} \label{subsec:2A}
We will consider a quantum system with a single bound state
$|\Phi_{\text{b}}\rangle$ with energy $-\chi$ and a continuum of free states $|\Phi_{\omega}\rangle$ indexed by the energy $\omega > 0$ with a density of states $\beta(\omega)$. This choice on the one hand describes well the intraband transitions in doped quantum wells in which photon-bound excitons have been predicted and observed for the first time \cite{cortese_strong_2019, cortese_excitons_2021}, on the other hand it is instantiated by the simplest mathematical model of a system presenting a ionising transition: a particle in a 1D space in the presence of an attractive Dirac delta potential. While we will keep our theoretical developments as general as possible, we will refer to such an iconic toy model, explicitly solved in Appendix \ref{App:A}, when required to make numerical examples. To simplify the notation in this paper we will work in units such that $\hbar=1$. 

Measuring energies from the bound state $|\Phi_{\text{b}}\rangle$, the Hamiltonian describing a ionizing electronic transition driven by a monochromatic laser field of frequency $\omega_0$ and field amplitude $E_0$, can be expressed in the rotating frame (Fig.~\ref{Fig:1}, panel (a)) as 
\begin{eqnarray}
H_{\text{S}}&=&\int_0^\infty d\omega  (\omega+\chi-\omega_0) \sigma_{\omega, \text{b}} \sigma_{\text{b},\omega} \nonumber\\&&+
\int_0^\infty d\omega   g(\omega)\left  [  \sigma_{\omega, \text{b}}+   \sigma_{\text{b},\omega} \right].
\label{HS}
\end{eqnarray}
Here, the transition operators are defined as  
$\sigma_{\text{b},\omega} = |\Phi_{\text{b}}\rangle \langle\Phi_{\omega}|$ and the coupling function is given by  $g(\omega) = e E_0 D(\omega),$  where \( D(\omega) \) is the frequency-dependent dipole moment. 
It is important to note that in our model only the bound‐to‐continuum transitions are coupled to the semiclassical pump laser, as these transitions predominantly govern the dynamics. Meanwhile, any significant dipolar coupling among continuum states occurs only between nearly degenerate levels, which remain off resonance with the laser.

\subsection{Diagonalization of the isolated system}\label{subsec:2B}
The spectrum of Eq.~\ref{HS}, whose diagonalization is detailed in Appendix \ref{App:B}, has in general two different components: a discrete resonance $|\Psi_{\text{d}} \rangle$ with energy $\omega_{\text{d}}$, and a continuum  $|\Psi_{\bar{\omega}} \rangle$ with energies $\bar{\omega}$.

The discrete eigenfrequency $\omega_{\text{d}}$, whose eigenfunction can be expressed as
\begin{eqnarray} \label{1}
|\Psi_{\text{d}} \rangle= d_{\text{b}} |\Phi_{\text{b}} \rangle +  \int_0^\infty d\omega   \, \, d(\omega) |\Phi_{\omega} \rangle,
\end{eqnarray}
satisfies the eigenequation
\begin{eqnarray}\label{disp}
\left(\omega_0-\omega_{\text{d}}\right)=\int_0^\infty d\omega \frac{|g(\omega)|^2}{\left(\omega+\chi-\omega_{\text{d}}\right)}.
\end{eqnarray}
It is easy to prove that a real solution $\omega_{\text{d}}$ to Eq.~\ref{disp} is guaranteed to exist if 
\begin{eqnarray}
\left(\omega_0-\chi\right)<\int_0^\infty d\omega\frac{|g(\omega)|^2}{\omega},
\label{existence}
\end{eqnarray}
and such a condition is trivially true for $\omega_0<\chi$, that is when the laser frequency is not large enough to ionise the electronic system with a single photon transition. The discrete resonance can in this case be interpreted as an off-resonant dressing of the laser mode by the continuum. In the opposite case $\omega_0>\chi$, the discrete resonance exists only for large-enough values of the light-matter coupling.
The continuum resonances $\bar{\omega}$, and their eigenstates 
\begin{eqnarray} \label{2}
|\Psi_{\bar{\omega}} \rangle= c_{\text{b}}(\bar{\omega}) |\Phi_{\text{b}} \rangle +  \int_0^\infty d\omega   \, \, c(\bar{\omega},\omega) |\Phi_{\omega} \rangle ,
\end{eqnarray}
can instead be found, using the Fano diagonalization procedure \cite{fano_effects_1961}, for all $\bar{\omega}\geq \chi$.
As derived in Appendix \ref{App:B}, the continuum eigenenergies can be expressed via a complex dispersion relation 
 \begin{eqnarray} \label{sing}
\bar{\omega}-\omega_0 -\Omega(\bar{\omega})+i G(\bar{\omega})=0,
 \end{eqnarray} 
 with effective collective coupling $
\Omega(\bar{\omega})=\mathcal{P}\int d\omega
 \frac{|g(\omega)|^2}{\left(\bar{\omega}-\omega'-\chi \right)}$, and effective loss rate $
 G(\bar{\omega})=\pi |g(\bar{\omega}-\chi)|^2$ due to the continuum leakage channel. These ionization modes are conceptually equivalent to Floquet quasienergy states satisfying Gamow–Siegert boundary conditions, each characterized by a single-photon, energy-dependent ionization rate. 

The spectra of the system in the two cases, using the 1D electronic model described in Appendix~\ref{App:A}, can be found in Fig.~\ref{Fig:2} (a-c). The spectra are plotted as a function of the normalised coupling strength 
\begin{eqnarray}
\label{eq:eta}
\eta=|g_{\text{max}}|^2/\chi,    
\end{eqnarray}
where $g_{\text{max}}=\max_{\omega}g(\omega)$ is the  maximum coupling strength over the continuum resonances;  in panel (a) the laser frequency is slightly below half of ionization threshold ($\omega_0=0.45 \chi $), in panel (b) below the ionization edge ($\omega_0=0.9\chi$)  and in panel (c)  above it, set resonant with the maximum dipole moment  $\omega_0=\max \arg_{\omega} g(\omega)$.  
Consistently with our prediction, when pumping occurs below the continuum ($\omega_0<\chi$), a discrete resonance with energy $\omega_{\text{d}}$ emerges at every coupling strength. In contrast, when the laser frequency lies above the ionization edge ($\omega_0>\chi$), a discrete mode appears only above a minimal coupling strength. These results closely mimic what observed in the quantum regime for photon-bound excitons. 
This is the first major result of this paper. In the following our objective will be to understand the impact of such a discrete resonance upon the system's dynamics and its resonance fluorescence spectrum.

\subsection{Coupling the system to the environment}
\label{subsec:2C}
In order to calculate the resonance fluorescence spectrum we have to pass to an open quantum system description, allowing transitions induced by spontaneous radiative emission between adjacent ladders of coupled resonances, characterised by different photon number (see panel (b) of Fig.~\ref{Fig:1}). Unlike previous studies where the continuum was treated as a passive uncolored loss channel and eventually adiabatically eliminated \cite{finkelstein-shapiro_nonlinear_2016,finkelstein-shapiro_adiabatic_2020}, we will have to explicitly model it as a set of harmonic oscillator baths at zero temperature, which act as loss channels for both energy and coherence.
Beyond the radiative transitions directly responsible for the fluorescence, we will also have to take care of non-radiative decay channels, which can be dominant in solid-state systems \cite{wagner_phonon_1990,wang_minimizing_2023,burdov_exciton-photon_2021}.
Following the theory developed in the Appendix~\ref{App:C} we can trace over the degrees of freedom of the reservoir and recover, under standard assumptions, a master equation for the density operator of the system $\rho_{\text{S}}$ in Lindblad form 
\begin{eqnarray} \label{Lrho}
\frac{d\rho_{\text{S}}(t)}{dt}=-i\left[H_{\text{S}}+H_{\text{LS}},\rho_{\text{S}}(t) \right]+\mathcal{L}\rho_{\text{S}}(t).
\end{eqnarray}
In Eq.~\ref{Lrho} \( \mathcal{L} \) is the Lindblad superoperator accounting for dissipation and $H_{\text{LS}}$ the Lamb shift Hamiltonian accounting for frequency renormalisation induced by the dissipation.

\subsection{Population dynamics} \label{sec:3}
To calculate the fluorescence of the system under continuous pumping, the first step is to solve the master equation in Eq.~\ref{Lrho} for the populations of the different resonances. As detailed in Appendix~\ref{App:C}, under appropriate approximations this leads to a system of rate equations for the populations in the discrete ($N_{\text{d}}$) and continuum $(N_{\bar{\omega}})$ resonances, linked by the sum rule of the density matrix trace 
\begin{eqnarray}
\label{trace equation}
    N_{\text{d}}+\int_{\chi}^{\infty} d\bar{\omega} N_{\bar{\omega}}=1.
\end{eqnarray}
We now introduce the radiative $\gamma^{\text{r}}(\omega_{\text{d}},\bar{\omega})$ and non-radiative $\gamma^{\text{nr}}(\omega_{\text{d}},\bar{\omega})$ transition rates between the discrete and continuum resonances. These rates originate from different coupling mechanisms and exhibit distinct spectral dependencies (as detailed in App.~\ref{App:C}), together defining the total transition rate
\begin{eqnarray}
 {\gamma}(\omega_{\text{d}},\bar{\omega})&=&  {\gamma}^{\text{r}}(\omega_{\text{d}},\bar{\omega})+{\gamma}^{\text{nr}}(\omega_{\text{d}},\bar{\omega}).
\end{eqnarray}
The rate equations read
\begin{eqnarray} \label{rate equation}
\dot{N}_{\text{d}}&=&- \Gamma_{\text{d}} N_{\text{d}},\\ 
\dot{N}_{\bar{\omega}}&=&{\gamma}(\omega_{\text{d}},\bar{\omega}) N_{\text{d}}, \nonumber
\end{eqnarray}
where
\begin{eqnarray}
\Gamma_{\text{d}}&=&\int_\chi^\infty d\bar{\omega}\,{\gamma}(\omega_{\text{d}},\bar{\omega}),
\end{eqnarray}
represents the total decay rate from the discrete resonance into the continuum.
From Eq.~\ref{rate equation} we can see that the population of the discrete resonance decays exponentially into the continuum resonances. This is the second major result of this paper: the discrete resonance created by the coupling with the coherent pump is metastable. The result is best understood in analogy to the metastable nature of photon-bound states created by single photons in a cavity  \cite{cortese_strong_2019,cortese_excitons_2021}, although the physical decay channel is different.
In the single photon case
 the lifetime is due to the photon escaping the cavity, while in the quasi-classical case the lifetime is due to two-photon ionization, or more correctly photon-polariton ionization, due to the excitation of a polaritonic state of frequency $\omega_{\text{d}}$ plus the emission of a red-detuned photon from the pump beam.
 
The rate equation can be solved with the system initially in its ground state when the pump is switched on, $\rho_{\text{S}}(0)=\lvert \Phi_{\text{b}} \rangle\langle \Phi_{\text{b}} \rvert$, leading to the initial condition $N_{\text{d}}(0)=\lvert d_{\text{b}}\rvert^2$.
 In Fig.~\ref{Fig:3} (d-i) we plot in each panel both the population dynamics of the discrete state $N_{\text{d}}(t)$ as a function of normalised time on the left, and the decay rates from the discrete to the continuum $\gamma(\omega_{\text{d}},\bar{\omega})$ on the right, for the three pumping scenarios depicted above and for different values of the normalised coupling $\eta$ from Eq.~\ref{eq:eta}.

 Because the population of the discrete resonance depends on the combined radiative, ${\gamma}^{\text{r}}(\omega_{\text{d}},\bar{\omega})$, and non-radiative, ${\gamma}^{\text{nr}}(\omega_{\text{d}},\bar{\omega})$, transition rates into the continuum, which differ substantially in both spectral dependence and magnitude in real physical systems,  we analyse two limiting cases:\\
 (i) The case of almost purely radiative losses, in which the total dissipation rate is much smaller than the binding energy 
 ($1\gg{\gamma}^{\text{r}}(\omega_{\text{d}},\bar{\omega})\gg {\gamma}^{\text{nr}}(\omega_{\text{d}},\bar{\omega})$ ), a situation typical of laser-driven atoms or simple molecules (illustrated in panels d-f). \\
(ii) A regime in which non-radiative losses dominate, with rates non-negligible compared to the binding energy, ($0.1\approx{\gamma}^{\text{nr}}(\omega_{\text{d}},\bar{\omega})\gg{\gamma}^{\text{r}}(\omega_{\text{d}},\bar{\omega})$), representative of solid-state systems with phonon-induced dissipation (shown in panels g-i). This regime describes well samples similar to the one used in the observation of PBE \cite{cortese_excitons_2021}: a stack of multiple highly doped GaAs quantum wells, under mid-infrared pumping quasi-resonant to the ionization frequency.

When  $\omega_0+\omega_{\text{d}}<\chi$ (a,d,g), the discrete eigenstate $\omega_{\text{d}}$ plus the energy of one photon $\omega_0$ can never exceed the continuum edge $\chi$, thus  two-photon, or more properly photon-polariton ionisation, becomes energetically forbidden and the populations remain stationary. 
Given that $\omega_{\text{d}}<\omega_0$, the condition $\omega_0<\chi/2$ ensures that no ionization is possible regardless of the light-matter coupling strength. 
When $\omega_{0}>\chi/2$, the discrete eigenmode can {\it a priori} decay into the continuum, and its decay rate depends on how strongly it hybridizes with that continuum. In the scenario $\chi>\omega_{0}>\chi/2$, increasing coupling strengthens this hybridization and thus increases the emission rate. By contrast, in the $\omega_0>\chi$ case, the mode initially lies within the continuum at weak coupling but becomes more discrete as coupling grows, which suppresses its emission into the continuum. 
As anticipated, differences between the impact of radiative or non-radiative losses depend also on the different density of states for the free-space photons responsible for radiative losses  (${\gamma}^{\text{r}}(\omega_{\text{d}},\bar{\omega})\propto(\omega_{\text{d}}+\omega_0-\bar{\omega})^2 $) and for the non-radiative decay channel  which is instead assumed to be white (${\gamma}^{\text{nr}}(\omega_{\text{d}},\bar{\omega})\propto\bar{\omega}^0$). 

\subsection{Resonance fluorescence spectrum}\label{sec:4}
The resonance fluorescence spectrum, according to the Wienier-Khinchin theorem, can be expressed as the real part of the Fourier transform of the correlation function of the dipole polarization operator
\begin{eqnarray} \label{S}
    S(\omega_{\text{f}})&=&\text{Re} \left [ \int_0^\infty  \text{lim}_{t\rightarrow\infty} \langle \sigma^+(t+\tau)  \sigma^-(t) \rangle e^{-i(\omega_{\text{f}}-\omega_0)\tau }d\tau\right],\nonumber\\
  \end{eqnarray}
with the collective dipole polarization operator 
\begin{eqnarray}
    \sigma^+= \int_0^\infty d\omega D(\omega) \sigma_{\omega, \text{b}},
\end{eqnarray}    
and evaluated by applying the Quantum Regression theorem \cite{scully_quantum_1999,del_valle_regimes_2011}.
The calculations, developed in the Appendix \ref{appD},
show how we arrive to the analytic expression of the fluorescence spectrum 
\begin{eqnarray} \label{Sdec}
    S(\omega_{\text{f}})\!&=&  \!S_{\text{d-d}}(\omega_{\text{f}} )+\!S_{\text{d-c}}(\omega_{\text{f}} ),
\end{eqnarray} 
where the two terms represent the contributions from transitions between different resonances due to the spontaneous emission of a photon with energy $\omega_{\text{f}}$: from discrete to discrete, $S_{\text{d-d}}(\omega_{\text{f}})$, and  from discrete to continuum $S_{\text{d-c}}(\omega_{\text{f}})$.  The associated decay channels are depicted in panel (i) of Figs.~\ref{Fig:4} and \ref{Fig:5}. Detailed calculations in App.~\ref{appD} further show that the fluorescence arising from continuum states vanishes in the continuum limit.

The balance between these two contributions is primarily dictated by how localised the emergent discrete level is.  When the discrete state \(\Psi_{\text{d}}\) is more tightly localised, occurring at lower couplings for \(\omega_{0}<\chi\) and at higher couplings for \(\omega_{0}>\chi\), the discrete–to–discrete term
\(S_{\text{d-d}}(\omega_{\text{f}})\) dominates; conversely, when \(\Psi_{\text{d}}\) is
more delocalised, the discrete–to–continuum contribution \(S_{\text{d-c}}(\omega_{\text{f}})\) prevails.
Figures \ref{Fig:4} and \ref{Fig:5} show the fluorescence spectra \(S(\omega_{\text{f}})\) of a laser-driven system in the same two limiting cases considered in the previous section: one with a weak, almost purely radiative decay channel  ($1\gg{\gamma}^{\text{r}}(\omega_{\text{d}},\bar{\omega})\gg {\gamma}^{\text{nr}}(\omega_{\text{d}},\bar{\omega})$, Fig.~\ref{Fig:4}) and one with large non-radiative dissipation ($\gamma^{\text{r}}(\omega_{\text{d}}, \bar{\omega})\ll\gamma^{\text{nr}}(\omega_{\text{d}}, \bar{\omega})\approx 0.1$, Fig.~\ref{Fig:5}).  
In both Figures, panels (a-c) depict the below two-photon threshold $\omega_0<\chi/2$ pumping regime, while panels (d-f) represent the below threshold conditions $\chi/2<\omega_0<\chi$, and panels (g-i)  the case where \( \omega_0 \) exceeds the ionization threshold \(\chi\) and aligns with the maximum of the coupling. The spectra are plotted for three different normalised coupling: \(\eta = 0.05\) (a,d,g), \(\eta = 0.1\) (b,e,h), and \(\eta = 0.2\) (c,f,i).
In each panel, the total fluorescence is plotted with a black solid line, and its two constituents from different terms of Eq.~\ref{Sdec} as solid lines, color coded according to their origin as shown in panel (i). 
In both Figures,  the first row (a-c) shows the trivial result of no resolved emission resonance for the below two-photon threshold condition $\omega_0<\chi/2$: no emission channel from the discrete state is allowed. 
The case depicted in Fig.~\ref{Fig:4} is characterised by long dissipation lifetime  relative to the photo-excitation dynamics; as a result, the discrete-to-continuum spectral component \(S_{\text{d-c}}(\omega_{\text{f}})\) (red curve) appears as the envelope of the emission spectra for all continuum states, with an intensity proportional to their degree of hybridisation with the bound state, which increases with coupling strength when $\omega_{0}<\chi$ and decreases with coupling when $\omega_{0}>\chi$. The value of the coupling where the hybridization $|c_{\text{b}}(\bar{\omega})|^2$ reaches its maximum is marked by a vertical red dotted line.
The elastic spectral component \(S_{\text{d-d}}(\omega_{\text{f}})\) (blue curve) represents the emission from the discrete state to itself. Being in this scenario its linewidth much smaller than other energy scales of the problem, it manifests as a sharp, delta-like resonance.

In Fig.~\ref{Fig:5} we display the fluorescence spectrum for a system whose dressed discrete‐state dynamics is dominated by non-radiative dissipation comparable to other energy scales of the problem. 
For $\chi/2<\omega_{0}<\chi$, the elastic peak dominates over the discrete-to-continuum contribution because the discrete state remains strongly localised throughout the entire coupling range.
In the above-threshold regime ($\omega_{0}>\chi$), the discrete state contains a substantial free-electron component, so it relaxes with comparable probability into itself and into the continuum.

\section{Conclusions}
\label{sec:5} 
In this paper, we investigated the laser-driven dynamics of an electronic system consisting of a single bound state and a continuum of free electronic states. Our semi-analytical open quantum system theory allowed us to demonstrate that new stable electronic states due to the resonant interaction with the laser exist, although they are metastable due to photon-polariton ionization.
We calculated the resulting fluorescence spectrum, which in the bound-to-bound transition case would lead to the well-known Mollow triplet. In our bound-to-continuum case, however, only two of the four emission channels remain, and they assume very different spectral features.
We showed how different kinds of spectra can be expected based on the dominant loss channel of the system and its scale in relation with the binding energy. Our results demonstrate a novel approach to measure stabilised electronic states in atomic, molecular or solid-state systems, creating stable electronic states by resonant pumping.  We investigated numerically two regimes. The first is representative of systems in which the PBEs have been experimentally observed, namely a stack of narrow (4-nm) and highly doped ($5\times 10^{12}$ cm$^{-2}$) GaAs quantum wells, embedded in a resonant mid-infrared resonator quasi resonant to the ionization frequency ($\approx140$ meV) \cite{cortese_excitons_2021}. The corresponding AC-Stark shift has been observed in similar intersubband devices using mid-infrared  laser with power of 5 MW/m$^2$ \cite{dynes2005ac}.
The second regime employs parameters more pertaining to atomic and molecular systems. In this context, several works over the past decades have reported stabilisation against multiphoton dissociation in hydrogen-like ions (H$_2^+$, D$_2^+$ \ldots.) using multiphoton Floquet channels \cite{dunn1968photodissociation, aubanel1993molecular}. More recently, an analogous form of stabilisation against dissociation has been observed in Rydberg macromers \cite{hollerith2024rydberg}, occurring in a regime where the vibrational states forming the continuum have an energy comparable to the binding energy. This observation suggests a similar mechanisms could also emerge in strongly driven diatomic molecular ions such as K$_2^+$, which possess a single discrete bound state resonantly coupled to a dissociating continuum, and whose dissociation energy is optically accessible from the ground state.
\begin{backmatter}
\bmsection{Funding}

\bmsection{Acknowledgment}
The authors acknowledge financial support from the Leverhulme Trust, grant No. RPG-2022-037. EC acknowledges financial support from the Engineering and Physical Sciences Research Council (EPSRC), grant No. EP/Y021835/1. 
SDL acknowledges financial support under the National Recovery and Resilience Plan (NRRP), Mission 4, Component 2, Investment 1.1, Call for tender No. 104 published on 02/02/2022 by the Italian Ministry of University and Research (MUR), funded by the European Union – NextGenerationEU – Project Title PENNA - CUP B53D23003790006 - Grant Assignment Decree No. 957 adopted on 30/06/2023 by the Italian Ministry of University and Research (MUR).

\bmsection{Disclosures}
The authors declare no conflicts of interest.

\bmsection{Data Availability Statement}
The data that support the findings of this study are available from the corresponding authors on reasonable request.

\end{backmatter}
\bibliography{references}

\appendix 
\section{Light-matter coupling in a 1D delta potential}\label{App:A}
In this Appendix we solve the problem of a 1D massive particle in presence of a Dirac-delta attractive potential, calculating the dipole moment between the bound and the free states.
We look for the eigensolutions of the time-independent Schroedinger equation
\begin{eqnarray}\label{SE}
- \frac{1}{2 m} \frac{d^2 \Phi(x)}{d x^2} - V \delta(x) \Phi(x)= E \Phi(x),
\end{eqnarray}
where $m$ is the mass of the particle and $V$ the potential's depth. 
Imposing continuity of the wavefunction at the origin we find the wavefunction of the single bound state 
\begin{eqnarray}
\Phi_{\text{b}}(x)=\sqrt{m V} e^{-V m |x|},
\end{eqnarray}
with negative energy $E=-\chi=-\frac{V^2 m}{2}$.
Being such a wavefunction even, only odd solutions are dipolarly coupled to it, but such solutions have to vanish at $x=0$ and they do not see the delta potential. The only other solutions we have to consider are thus free particle solutions 
\begin{eqnarray}
\Phi_{k}(x)=\sqrt{\frac{2}{L}}\sin(k x),
\end{eqnarray}
with positive energy $\omega=\frac{k^2}{2m}>0$, $k$ the wavevector, and $L$ the quantization length corresponding to a  density of states in the continuum $\beta(\omega)=\frac{L}{\pi}\sqrt{\frac{m}{2\omega}}$. The gap of the system, corresponding to the first ionization energy is thus equal to $\chi$.
The wavevector-dependent dipole moment of the ionising transition can thus be written as
\begin{eqnarray}
D_k&=&\int_{-\infty}^\infty \Phi^{*}_k(x) \,x \,\Phi_{\text{b}}(x) dx 
= \sqrt{\frac{2m V}{L}} \int_{-\infty}^\infty \sin(kx) \,x\, e^{-V m |x|} dx =
\sqrt{\frac{8}{m\pi\beta(\omega)}}\frac{\chi^{3/4}\omega^{1/4}}{(\chi+\omega)^2},
\end{eqnarray}
and the energy-dependent one used in the main body of the text is thus given by
\begin{eqnarray} \label{Dw}
D(\omega)&=&D_k\sqrt{\beta(\omega)}=
\sqrt{\frac{8}{m\pi}}\frac{\chi^{3/4}\omega^{1/4}}{(\chi+\omega)^2}.
\end{eqnarray}

\section{Diagonalization of the laser-dressed system}\label{App:B}
We consider a monochromatic polarised laser field of frequency $\omega_0$ coupled to a set of ionizing electronic transitions between the single bound state $| \Phi_{\text{b}} \rangle$ with energy $-\chi$ and the free states $| \Phi_\omega \rangle$ with positive energies $\omega$. 
Choosing the energy of the bound state as reference and working in the Rotating Wave Approximation (RWA) the resulting Hamiltonian can be written as 
\begin{eqnarray} 
\label{time}
{H}_{\text{S}}(t)&=&\int_0^\infty d\omega (\omega+\chi) \sigma_{\omega, \text{b}} \sigma_{\text{b},\omega} +\int_0^\infty d\omega g(\omega) \left [ e^{i\omega_0 t} \sigma_{\omega, \text{b}}+e^{-i\omega_0 t} \sigma_{\text{b},\omega}\right],
\end{eqnarray}
where the $\sigma_{\text{b},\omega}=| \Phi_{\text{b}} \rangle\langle \Phi_{\omega}| $ are the ionizing transition operators.
The time-dependency of the Hamiltonian in Eq.~\ref{time} can be removed passing in the rotating frame via the transformation 
\begin{eqnarray}
{U}(t)=e^{i \omega_0 t  \int d\omega \sigma_{\omega,\text{b}}\sigma_{\text{b},\omega}},
\end{eqnarray}
leading to time-independent Hamiltonian 
\begin{eqnarray}
{H}_{\text{S}}&=& \int_0^\infty d\omega  (\omega+\chi-\omega_0) \sigma_{\omega,\text{b}} \sigma_{\text{b},\omega}  \\
&&+\int_0^\infty d\omega  g(\omega)\left  [  \sigma_{\omega, \text{b}}+   \sigma_{\text{b},\omega} \right].\nonumber
\end{eqnarray}
The spectrum of this Hamiltonian has generally both a discrete and a continuum part \cite{passante_bound_2000}. We start from considering a discrete eigenfrequency $\omega_{\text{d}}$, whose eigenvector can be expressed as
\begin{eqnarray} \label{A1}
|\Psi_{\text{d}} \rangle= d_{\text{b}} |\Phi_{\text{b}} \rangle +  \int_0^\infty d\omega   \, \, d(\omega) |\Phi_{\omega} \rangle.
\end{eqnarray}
The resulting eigensystem reads
\begin{eqnarray} \label{coeff}
\left(\omega_{\text{d}}-\omega_0\right) d_{\text{b}}&=& \int_0^\infty d\omega  g(\omega) d(\omega),\nonumber \\ \label{e}
\left(\omega_{\text{d}}-\omega-\chi \right) d(\omega)&=& g(\omega) d_{\text{b}},
\end{eqnarray}
leading to the integral eigenequation
\begin{eqnarray}\label{secular}
\left(\omega_{\text{d}}-\omega_0\right)=\int_0^\infty d\omega\frac{|g(\omega)|^2}{\left(\omega_{\text{d}}-\omega-\chi \right)}.
\end{eqnarray}

To avoid confusion, we emphasise that $\omega_\textrm{d}$ is not a specific value of the shifted continuum frequency $\omega+\chi$, but it labels the single discrete hybrid mode that appears below the continuum. 
The condition for the existence of a discrete resonance, $\omega_{\text{d}} < \chi$, is 
\begin{eqnarray}
\left(\omega_0 - \chi\right) < \int_0^\infty d\omega\, \frac{|g(\omega)|^2}{\omega},
\label{existence}
\end{eqnarray}
and, for the specific toy model considered here, inserting the expression for the dipolar coupling given in Eq.~(\ref{Dw}) into Eq.~(\ref{existence}) and solving for the electric field amplitude yields

\begin{equation}
  E_0> \frac{\hbar \chi}{e} \sqrt{\frac{2m(\omega_0-\chi)}{5}}.  
\end{equation}

Supposing that Eq.~\ref{secular} has at least a real solution, by imposing the normalization condition $\langle \Psi_{\text{d}}|\Psi_{\text{d}} \rangle=1 $, we arrive to the following expressions for the coefficients 
\begin{eqnarray}   \label{eqdb}
d_{\text{b}}&=&\left[1+\int_0^\infty d\omega\frac{|g(\omega)|^2}{(\omega_{\text{d}}-\omega-\chi)^2}\right]^{-1/2},\!\\
    d(\omega)&=&\!\frac{g(\omega)}{\omega_{\text{d}}-\omega-\chi} \left[1+\int_0^\infty d\omega' \frac{|g(\omega')|^2}{(\omega_{\text{d}}-\omega'-\chi)^2}\right]^{-1/2}.\nonumber
\end{eqnarray}

For the continuum part of the spectrum, with eigenfrequencies $\bar{\omega}$, we write instead 
\begin{eqnarray}\label{A2}
 |\Psi_{\bar{\omega}} \rangle= c_{\text{b}}(\bar{\omega}) |\Phi_{\text{b}} \rangle +  \int_0^\infty d\omega   \, \, c(\omega,\bar{\omega}) |\Phi_{\omega} \rangle ,
\end{eqnarray}
with eigensystem
\begin{eqnarray} \label{unbound_coeff}
\left(\bar{\omega}-\omega_0\right) c_{\text{b}}(\bar{\omega})&=& \int_0^\infty d\omega  g(\omega) c(\omega,\bar{\omega}),\nonumber \\ \label{e2}
\left(\bar{\omega}-\omega-\chi \right) c(\omega,\bar{\omega})&=& g(\omega) c_{\text{b}}(\bar{\omega}).
\end{eqnarray}
Contrary to the case of a discrete eigenvalue, we can not solve this system simply dividing the second equation by $(\bar{\omega}-\omega-\chi)$, as there will always be a solution $\bar{\omega}$ which makes such an expression vanish. Following Fano \cite{fano_effects_1961} we thus introduce an unknown
function $y(\bar{\omega})$ to regularise the resulting pole
\begin{eqnarray} \label{unbound_coeff}
&&c(\omega,\bar{\omega})= \left[\mathcal{P}\frac{1}{\left(\bar{\omega}-\omega-\chi \right) }+y(\bar{\omega}) \delta\left(\bar{\omega}-\omega-\chi \right)\right]g(\omega) c_{\text{b}}(\bar{\omega}),\nonumber 
\\
&&\left(\bar{\omega}-\omega_0\right) c_{\text{b}}(\bar{\omega})= \int_0^\infty d\omega  g(\omega) c(\omega,\bar{\omega}),\label{e3}
\end{eqnarray}
leading to the equation
 \begin{eqnarray}
 \left(\bar{\omega}-\omega_0\right)&=&\mathcal{P}\int d\omega \frac{|g(\omega)|^2}{\left(\bar{\omega}-\omega-\chi \right)}+|g(\bar{\omega}-\chi)|^2y(\bar{\omega}),\nonumber\\
 \end{eqnarray} 
which should not be interpreted as an equation fixing the eigenvalues of the system, given that in the continuum all frequencies $\bar{\omega}$ are eigenfrequencies, but as an equation allowing to determine $y(\bar{\omega})$.
Finally, imposing the normalization of the continuum resonances
 \begin{eqnarray}
 \label{eqy}
 \langle \Psi_{\bar{\omega}}| \Psi_{\bar{\omega}'} \rangle=\delta(\bar{\omega}-\bar{\omega}'),
 \end{eqnarray} 
 we obtain
 \begin{eqnarray}
\label{eqcg}
 c_{\text{b}}(\bar{\omega})&=&\frac{1}{g(\bar{\omega}-\chi)} \frac{1}{\sqrt{y(\bar{\omega})^2+\pi^2}},
 \end{eqnarray} 
 where we picked the phase factor which makes $c_{\text{b}}(\bar{\omega})$ real. The lengthy expression of $c(\omega,\bar{\omega})$ can at this point be trivially calculated by solving Eq.~\ref{eqy} for $y(\bar{\omega})$ and plugging the resulting expression from Eq.~\ref{eqcg} into Eq.~\ref{unbound_coeff}.  The complex dispersion relation for the continuum dressed modes can be then extracted as pole of the coefficient $c_{\text{b}}(\bar{\omega})$, as 
 \begin{eqnarray} \label{sing}
\bar{\omega}-\omega_0 -\Omega(\bar{\omega})+i G(\bar{\omega})=0,
 \end{eqnarray} 
 with 
 \begin{eqnarray}
\Omega(\bar{\omega})&=&\mathcal{P}\int d\omega 
 \frac{|g(\omega)|^2}{\left(\bar{\omega}-\omega-\chi \right)},\\
 G(\bar{\omega})&=&\pi |g(\bar{\omega}-\chi)|^2.
 \end{eqnarray}
Such dispersion includes an effective collective coupling $\Omega(\bar{\omega})$ and an effective loss rate $G(\bar{\omega})$ due to the continuum leakage channel.

\section{Derivation of master equation in the Lindblad form}\label{App:C}
 In order to model how the system described by the Hamiltonian $H_{\text{S}}$ can interact with the environment , we start by introducing the Hamiltonians $H_{\text{R}}$ and $H_{\text{I}}$, describing respectively a reservoir able to exchange energy and coherence with the system, and the coupling between the system and the reservoir which describes such exchanges. 

The evolution of the total density matrix $\rho$ describing both system's and reservoir's degrees of freedom then follows the Heisenberg equation 
\begin{eqnarray}
\frac{d\rho}{dt}=-i\left[H_{\text{S}}+ H_{\text{R}}+H_{\text{I}}, \rho\right].
\end{eqnarray}
\begin{eqnarray}
H_{\text{R}}=\sum_\lambda \omega_\lambda \xi^\dagger_\lambda \xi_\lambda+\sum_\lambda \omega_\lambda \alpha^\dagger_\lambda \alpha_\lambda,
\end{eqnarray}
with $\xi_\lambda$ the bosonic annihilation operators for the matter reservoir (e.g., phonons) and $\alpha_\lambda$ the bosonic annihilation operator for emitted free photons. The interaction Hamiltonian describes linear couplings between the quadrature of the reservoirs and an electronic transition in the system, thus taking the general form
$H_{\text{I}}=H^{\text{r}}_{\text{I}}+H_{\text{I}}^{\text{nr}}$ , where
\begin{eqnarray} \label{HIr}
{H}^{\text{r}}_{\text{I}}(t)&=& \sum_\lambda \left[\int_0^\infty d\omega \mu_\lambda (\omega) {\Sigma}(\omega,t) R_\lambda(t)\right],\\ \label{HIn}
{H}^{\text{nr}}_{\text{I}}(t)&=& \sum_\lambda \left[\int_0^\infty d\omega \nu(\omega) {\Sigma}(\omega,t) T_\lambda(t)\right],\nonumber
\end{eqnarray}
with operators
\begin{eqnarray}
 {\Sigma}(\omega,t)&=& e^{-i \omega_0 t} e^{i H_{\text{S}} t}\sigma_{\text{b},\omega}  e^{-i H_{\text{S}} t}+\text{h.c.}, \\
{R}_\lambda(t)&=& \alpha_\lambda e^{-i \omega_\lambda t}+ \alpha^\dagger_\lambda e^{i \omega_\lambda t},\nonumber \\ 
 {T}_\lambda(t)&=& \xi_\lambda e^{-i \omega_\lambda t}+ \xi^\dagger_\lambda e^{i \omega_\lambda t}. \nonumber
\end{eqnarray}
Here $R_\lambda$ is a quadrature of the photonic bath in which the system is emitting photons, and $T_\lambda$ a quadrature of the non-radiative bath where the excitations is lost without emitting a photon. 
In Eqs.~\ref{HIr}-\ref{HIn} we specialise the radiative couplings by the expression $\mu_\lambda(\omega)=\mu\sqrt{\omega_\lambda} D(\omega)$, corresponding to the physical electric dipole coupling, and the non-radiative coupling by the phenomenological expression
$\nu(\omega)=\nu\sqrt{\beta(\omega)}$,  where $\beta(\omega)$ is the density of electronic states in the continuum.  
Following the general theory from Ref.~\cite{manzano_short_2020}, under the assumption of a Markovian evolution and applying the Rotating Wave Approximation (RWA) to eliminate rapidly oscillating terms, we derive a master equation in Lindblad form describing the evolution of the system density operator $\rho_{\text{S}} = Tr_{\text{R}}\left[\rho \right]$ 
\begin{eqnarray} \label{CLrho_app}
\frac{d\rho_{\text{S}}(t)}{dt}=-i\left[H_{\text{S}}+H_{\text{LS}},\rho_{\text{S}}(t) \right]+\mathcal{L}\rho_{\text{S}}(t),
\end{eqnarray}
where $\mathcal{L}=\mathcal{L}^{\text{r}}+\mathcal{L}^{\text{nr}}$ is the Lindblad superoperator  accounting for radiative 
(\( \mathcal{L}^{\text{r}} \))  as well as non-radiative  (\( \mathcal{L}^{\text{nr}} \))  dissipation. $H_{\text{LS}}$ denotes the Lamb-shift Hamiltonian arising from the coupling to the bath.  It is diagonal in the basis of the coupled resonances defined by the projectors
\begin{eqnarray}
P_{\text{d},\text{d}} &=& |\Psi_{\text{d}} \rangle \langle \Psi_{\text{d}}|,\\
P_{\bar{\omega},\text{d}} &=& |\Psi_{\bar{\omega}} \rangle \langle \Psi_{\text{d}}|,\\
P_{\bar{\omega},\bar{\omega}'} &=& |\Psi_{\bar{\omega}} \rangle \langle \Psi_{\bar{\omega}'}|.
\end{eqnarray}
The Lamb-shift Hamiltonian can then be written as
\begin{eqnarray}
H_{\text{LS}} = 
\tilde{\omega}_{\text{d}}\, P_{\text{d},\text{d}} 
+ \int d\bar{\omega}\, \tilde{\bar{\omega}}\, P_{\bar{\omega},\bar{\omega}},
\end{eqnarray}
where $\tilde{\omega}_{\text{d}}$ and $\tilde{\bar{\omega}}$ are the Lamb-shifted eigenenergies of the discrete and continuum-like resonances, respectively,
\begin{eqnarray}
\tilde{\omega}_{\text{d}} 
&=& \omega_{\text{d}} 
+  \sum_{i=\text{r},\text{nr}}\left[
\pi^{i}(\omega_{\text{d}},\omega_{\text{d}})
+ \int d\omega\, \pi^{i}(\omega_{\text{d}},\omega)
\right], \\
\tilde{\bar{\omega}} 
&=& \bar{\omega} 
+ \sum_{i=\text{r},\text{nr}}\left[
\beta^{-1}(\bar{\omega})\, \pi^{i}(\bar{\omega},\omega_{\text{d}})
+ \int d\omega\, \beta^{-1}(\bar{\omega})\, \pi^{i}(\bar{\omega},\omega)
\right].
\end{eqnarray}
The functions $\pi^{i}(\omega_1,\omega_2)$ ($i=\text{r},\text{nr}$) encode the radiative and non-radiative energy shifts and can be explicitly expressed as
\begin{eqnarray}
\pi^{\text{r}}(\omega_1,\omega_2)
&=&
\sum_{\lambda}\mu^2 \omega_\lambda
\Bigg[
x_{\text{b}}(\omega_1)^2 X^{\text{r}}(\omega_2)^2 
\mathcal{P}\!\left(\frac{1}{\omega_2 - \omega_1 + \omega_0 - \omega_\lambda}\right)
+ x_{\text{b}}(\omega_2)^2 X^{\text{r}}(\omega_1)^2 
\mathcal{P}\!\left(\frac{1}{\omega_2 - \omega_1 - \omega_0 - \omega_\lambda}\right)
\Bigg], \nonumber\\ \\
\pi^{\text{nr}}(\omega_1,\omega_2)
&=&
\sum_{\lambda}\nu^2
\Bigg[
x_{\text{b}}(\omega_1)^2 X^{\text{nr}}(\omega_2)^2 
\mathcal{P}\!\left(\frac{1}{\omega_2 - \omega_1 + \omega_0 - \omega_\lambda}\right)
+ x_{\text{b}}(\omega_2)^2 X^{\text{nr}}(\omega_1)^2 
\mathcal{P}\!\left(\frac{1}{\omega_2 - \omega_1 - \omega_0 - \omega_\lambda}\right)
\Bigg].\nonumber\\ 
\end{eqnarray} 
The coefficients $x_{\text{b}}(\omega)$ and $X^{i}(\omega)$ ($i=\text{r},\text{nr}$) describe the projection of the dressed states onto the bare basis and depend on whether the resonance is discrete or embedded in the continuum:
\begin{eqnarray}  \label{Lambda}
\Lambda^{\text{r}}_{\text{d}} &=& \int_0^\infty d\omega\, D(\omega)\, d(\omega), \\
{\Lambda}^{\text{r}}(\bar{\omega}) &=& \int_0^\infty d\omega\, D(\omega)\, c(\omega,\bar{\omega}), \nonumber\\
\Lambda^{\text{nr}}_{\text{d}} &=& \int_0^\infty d\omega\, \sqrt{\beta(\omega)}\, d(\omega), \nonumber\\
{\Lambda}^{\text{nr}}(\bar{\omega}) &=& \int_0^\infty d\omega\, \sqrt{\beta(\omega)}\, c(\omega,\bar{\omega}). \nonumber
\end{eqnarray}
In particular, $x_{\text{b}}(\omega)$ and $X^{i}(\omega)$ take the values $\{d_{\text{b}}, \Lambda^{i}_{\text{d}}\}$ for a discrete resonance, and $\{c_{\text{b}}(\bar{\omega}), \Lambda^{i}(\bar{\omega})\}$ for a continuum resonance. In this paper, we considered that the physical frequencies are already renormalised and thus did not explicitly consider the effect of $H_{\text{LS}}$.

It is worth mentioning that in a general multi‐level system with nearly degenerate transition energies, employing a strict RWA may be inaccurate, because it discards terms oscillating on timescales comparable to the system’s relaxation dynamics. A more refined approach, often referred to as the partial secular approximation (PSA), accounts for all transitions whose oscillation frequencies are not too far apart compared to the relaxation time, thereby capturing additional subspace couplings \cite{cattaneo_symmetry_2020,trushechkin_unified_2021}.
In our continuum setting, however, going beyond exactly degenerate transitions would lead to contributions that, owing to the intrinsically incoherent and dissipative nature of continuum states, would have no physical effects beyond a potential slight change in the resonant lineshapes. We thus completely neglect non-degenerate transitions. 

The Lindblad superoperator can be expressed as ($i=\{\text{r}, \text{nr}\}$)
\begin{eqnarray}
\label{lcomp}
\mathcal{L}^i  = \mathcal{L}^i_{\text{d-d}}   + \mathcal{L}^i_{\text{d-c}}  + \mathcal{L}^i_{\text{c-c}}  + \mathcal{L}{^{i}}',
\end{eqnarray}
separating its distinct contributions: discrete-discrete (\( \mathcal{L}_{\text{d-d}} \)), discrete-continuum (\( \mathcal{L}_{\text{d-c}} \)), and continuum-continuum (\( \mathcal{L}_{\text{c-c}} \)) decay processes, along with the additional contribution $\mathcal{L}'$ that accounts for transitions that survive the RWA as involving equal transition energies but different resonance pairs. These terms, absent in discrete mode systems, emerge from the continuum nature of the coupled states, acting as a potential mechanism for coherence transfer between pairs of states.  
Assuming zero temperature, such that the equilibrium expectation value of $\xi_{\lambda}^\dagger \xi_{\lambda}$ and $\alpha_\lambda^\dagger \alpha_\lambda$ is always vanishing and the reservoir only acts as an energy sink, the superoperators in Eq.~\ref{lcomp} can be explicitly written as
\begin{eqnarray}
\label{Lexplicit}
\mathcal{L}^i_{\text{d-d}} \rho_{\text{S}} &=& \gamma^i({\omega}_{\text{d}},{\omega}_{\text{d}}) \left[ P_{\text{d},\text{d}} \rho_{\text{S}} P_{\text{d},\text{d}} - \frac{1}{2} \left\{ P_{\text{d},\text{d}} , \rho_{\text{S}}  \right\} \right],\\
\mathcal{L}^i_{\text{d-c}} \rho_{\text{S}} &=& \int d\bar{\omega} \beta^{-1}(\bar{\omega}) \gamma^i (\bar{\omega},{\omega}_{\text{d}}) \left[P_{d,\bar{\omega}} \rho_{\text{S}}  P_{\bar{\omega},\text{d}} - \frac{1}{2} \left\{ P_{\bar{\omega},\bar{\omega}},\rho_{\text{S}}\right\}\right]  \nonumber\\
&&  + \int d\bar{\omega} \beta^{-1}(\bar{\omega}) \gamma ^i({\omega}_{\text{d}},\bar{\omega}) \left[ P_{\bar{\omega},\text{d}} \rho_{\text{S}}  P_{d,\bar{\omega}} - \frac{1}{2} \beta(\bar{\omega}) \left\{ P_{\text{d},\text{d}},\rho_{\text{S}} \right\} \right ],\\ 
\mathcal{L}^i_{\text{c-c}}\rho_{\text{S}} &=& \int d\bar{\omega} \int d\bar{\omega}' \int d\bar{\omega}'' \beta^{-1}(\bar{\omega}'-\bar{\omega}+\bar{\omega}'') \gamma^i (\bar{\omega},\bar{\omega}',\bar{\omega}'-\bar{\omega}+\bar{\omega}'',\bar{\omega}'') \nonumber\\
&&  \times \left[P_{\bar{\omega},\bar{\omega}'}\rho_{\text{S}}  P_{\bar{\omega}''+\bar{\omega}'-\bar{\omega},\bar{\omega}''} - \frac{1}{2} \delta(\bar{\omega}''-\bar{\omega}) \left \{ P_{\bar{\omega}''+\bar{\omega}'-\bar{\omega},\bar{\omega}'},\rho_{\text{S}}\right\} \right],\\ \label{Lprime}
\mathcal{L}{^{i}}' \rho_{\text{S}} &=& \int d\bar{\omega} \beta^{-1} (\bar{\omega}) {\gamma} ^i(\bar{\omega},\bar{\omega},{\omega}_{\text{d}},{\omega}_{\text{d}}) \left(  P_{\text{d},\text{d}} \rho_{\text{S}}  P_{\bar{\omega},\bar{\omega}} + P_{\bar{\omega},\bar{\omega}} \rho_{\text{S}} P_{\text{d},\text{d}}  \right) \nonumber\\
&& + \int d\bar{\omega} \int d\bar{\omega}' \beta^{-1}(\bar{\omega}'-{\omega}_{\text{d}}+\bar{\omega}) {\gamma}^i ({\omega}_{\text{d}},\bar{\omega},\bar{\omega}'-{\omega}_{\text{d}}+\bar{\omega},\bar{\omega}') P_{\text{d},\bar{\omega}} \rho_{\text{S}} P_{\bar{\omega}'-{\omega}_{\text{d}}+\bar{\omega},\bar{\omega}'} \nonumber\\
&& + \int d\bar{\omega} \int d\bar{\omega}' \beta^{-1}({\omega}_{\text{d}}-\bar{\omega}+\bar{\omega}') {\gamma}^i(\bar{\omega},{\omega}_{\text{d}},\bar{\omega}'+{\omega}_{\text{d}}-\bar{\omega},\bar{\omega}') P_{\bar{\omega},\text{d}} \rho_{\text{S}} P_{\bar{\omega}'+{\omega}_{\text{d}}-\bar{\omega},\bar{\omega}'} \nonumber\\
&& + \int d\bar{\omega} \int d\bar{\omega}' \beta^{-1}({\omega}_{\text{d}}+\bar{\omega}-\bar{\omega}') {\gamma}^i(\bar{\omega},\bar{\omega}',{\omega}_{\text{d}},{\omega}_{\text{d}}-\bar{\omega}'+\bar{\omega}) P_{\bar{\omega},\bar{\omega}'} \rho_{\text{S}} P_{\text{d},{\omega}_{\text{d}}-\bar{\omega}'+\bar{\omega}}  \nonumber\\
&& + \int d\bar{\omega} \int d\bar{\omega}' \beta^{-1}({\omega}_{\text{d}}+\bar{\omega}-\bar{\omega}')  {\gamma}^i(\bar{\omega}',\bar{\omega},{\omega}_{\text{d}}-\bar{\omega}'+\bar{\omega},{\omega}_{\text{d}}) P_{\bar{\omega}',\bar{\omega}} \rho_{\text{S}} P_{{\omega}_{\text{d}}-\bar{\omega}'+\bar{\omega},\text{d}},
\end{eqnarray}
with $\left\{\cdot,\cdot \right\}$ the anticommutator.

The coefficients ${\gamma}^{i}(\omega_1,\omega_2,\omega_3,\omega_4)$ which appear
in Eqs.~\ref{Lexplicit}-\ref{Lprime} represent scattering coefficients from a resonance with frequency $\omega_1$ to one with frequency $\omega_2$, and from a resonance with frequency $\omega_4$ to one with frequency $\omega_3$, induced by the interaction with the reservoir, and they can be written as
\begin{eqnarray}  \label{gamma_ext}
{\gamma}^{\text{r}}(\omega_1,\omega_2,\omega_3,\omega_4)&=& \sum_\lambda 2{\mu}^2 \omega_\lambda x_{\text{b}}(\omega_1) x_{\text{b}}(\omega_4)X^{\text{r}}(\omega_2)X^{\text{r}}(\omega_3) \delta(\omega_1-\omega_2-\omega_0+\omega_\lambda)\\ 
&&+\sum_\lambda 2\mu^2 \omega_\lambda x_{\text{b}}(\omega_2) x_{\text{b}}(\omega_3)X^{\text{r}}(\omega_1)X^{\text{r}}(\omega_4) \delta(\omega_1-\omega_2+\omega_0+\omega_\lambda)\nonumber,\\
{\gamma}^{\text{nr}}(\omega_1,\omega_2,\omega_3,\omega_4)&=& \sum_\lambda 2{\nu}^2  x_{\text{b}}(\omega_1) x_{\text{b}}(\omega_4)X^{\text{nr}}(\omega_2)X^{\text{nr}}(\omega_3) \delta(\omega_1-\omega_2-\omega_0+\omega_\lambda)\nonumber \\ 
&&+\sum_\lambda 2\nu^2  x_{\text{b}}(\omega_2) x_{\text{b}}(\omega_3)X^{\text{nr}}(\omega_1)X^{\text{nr}}(\omega_4) \delta(\omega_1-\omega_2+\omega_0+\omega_\lambda)\nonumber.
  \end{eqnarray}

 Finally, the coefficients
 \begin{eqnarray}
 \gamma^{i}(\omega_1,\omega_2)&=&  {\gamma}^{i}(\omega_2,\omega_1,\omega_1,\omega_2),
 \end{eqnarray}
are contractions of the coefficients from Eq.~\ref{gamma_ext} and provide the scattering rate from a resonance with frequency $\omega_2$ to a resonance with frequency $\omega_1$ by emitting a photon. Although obtained by a more convoluted approach these coefficients are exactly the same that would result applying the standard Fermi Golden Rule.
Note that the function \({\gamma}(\omega_1,\omega_2)\) has been overloaded and its dimension depends upon its variables belonging to the discrete of continuum part of the spectrum. When both frequencies lie at the discrete resonance (\(\omega_1=\omega_2=\omega_{\text{d}}\)), \({\gamma}\) has the dimension of frequency. In contrast, for transitions between the discrete resonance and a continuum one (e.g. \(\omega_1=\omega_{\text{d}}\) and \(\omega_2=\bar{\omega}\)), it becomes dimensionless. Finally, when both frequencies belong to the continuum (\(\omega_1=\bar{\omega}\) and \(\omega_2=\bar{\omega}'\)), \({\gamma}\) takes the dimension of an inverse frequency. 
When multiple states possess closely spaced or identical Bohr frequencies, the terms associated with \(\mathcal{L}'\) in the Lindbladian introduce cross-dependence among different coherences and can induce coherence transfer between distinct decay channels. However, in our system the only source of coherence is the external driving that already dresses the bare states giving rise to the coupled resonances. Consequently, no net coherence can build up in the continuum, whose states rapidly dephase. As a result, and in line with previous studies on cross‐coherence interactions \cite{trushechkin_unified_2021}, we assume that the off‐diagonal elements of the density matrix decay quickly enough that they do not significantly influence the populations during the transient.
As such, we have neglected the cross terms arising from the additional \(\mathcal{L}'\) operator.

From the master equation in Eq.~\ref{CLrho_app} and the analytical expression of the Lindblad dissipator components in Eqs.~\ref{Lexplicit}-\ref{Lprime}, we can derive a set of rate equations for the populations and the off-diagonal coherences 

\begin{eqnarray}
\rho_{\text{d},\bar{\omega}}(t)&=&\langle\Psi_{\text{d}} | {\rho}_{\text{S}}(t)| \Psi_{\bar{\omega}}  \rangle,\quad \rho_{\bar{\omega},\text{d}}(t)=\langle\Psi_{\bar{\omega}} | {\rho}_{\text{S}}(t)| \Psi_{\text{d}}  \rangle, \quad \rho_{\bar{\omega},\bar{\omega}'}(t)=\langle\Psi_{\bar{\omega}} |{\rho}_{\text{S}}(t)| \Psi_{\bar{\omega}'} \rangle.
\end{eqnarray}
The full rate equations reads
    \begin{eqnarray} \label{re1}
\dot{N}_{\text{d}}&=&\sum_{i=\text{r,nr}}\left[\int_\chi^\infty d\bar{\omega}\beta^{-1}(\bar{\omega}) \gamma^{i}(\bar{\omega},\omega_{\text{d}}) N_{\bar{\omega}}\right]-\Gamma_{\text{d}}N_{\text{d}}, \\ 
\dot{N}_{\bar{\omega}}&=&\sum_{i=\text{r,nr}}\left[\gamma^{i}(\omega_{\text{d}},\bar{\omega}) N_{\text{d}}-\beta^{-1} (\bar{\omega})\gamma^{i}(\bar{\omega},\omega_{\text{d}}) N_{\bar{\omega}}-\beta^{-1} (\bar{\omega})\int_0^{\bar{\omega}+\omega_0} d\bar{\omega}' \gamma^{i}(\bar{\omega},\bar{\omega}') N_{\bar{\omega}}+\right.\nonumber\\
&&\left.\int_{\bar{\omega}-\omega_0}^\infty d\bar{\omega}' \beta^{-1} (\bar{\omega}')\gamma(\bar{\omega}',\bar{\omega}) N_{\bar{\omega}'}\right],\nonumber 
    \label{re3}\nonumber\\
    \dot{\rho}_{{\text{d}},\bar{\omega}}&=& -i (\omega_{\text{d}}-\bar{\omega})\rho_{{\text{d}},\bar{\omega}}+ \left[\rho_{{\text{d}},\bar{\omega}} \Gamma(\omega_{\text{d}},\bar{\omega})\right],\nonumber\\ \label{re4}
    \dot{\rho}_{\bar{\omega},{\text{d}}}&=& -i(\bar{\omega}-\omega_{\text{d}})\rho_{\bar{\omega},d}+\left[\rho_{\bar{\omega},{\text{d}}} \Gamma(\bar{\omega},\omega_{\text{d}})\right],\nonumber\\ \label{re5}
\dot{\rho}_{\bar{\omega},\bar{\omega}'}&=&-i(\bar{\omega}-\bar{\omega}')\rho_{\bar{\omega},\bar{\omega}'}+ \sum_{i=\text{r,nr}} \left[ \delta(\bar{\omega}-\bar{\omega}')\beta^{-1}(\bar{\omega}) \gamma^{i}(\omega_{\text{d}},\bar{\omega})N_{\text{d}}\right]+\Gamma(\bar{\omega},\bar{\omega}') \rho_{\bar{\omega},\bar{\omega}'},\nonumber
\end{eqnarray}
with effective decay rates
\begin{eqnarray}\label{rates}
\Gamma_{\text{d}}\!\!\!\!&=&\!\!\!\!   \sum_{i=\text{r,nr}}\int d\bar{\omega} \gamma^{i}(\omega_{\text{d}},\bar{\omega}),\\
 \Gamma(\omega_{\text{d}},\bar{\omega})\!\!\!\!&=&\!\!\!\! -\frac{1}{2}\sum_{i=\text{r,nr}} \left( \gamma^{i}(\omega_{\text{d}},\omega_{\text{d}})+  \int_\chi^\infty d\bar{\omega}' \beta^{-1}(\bar{\omega}')  \gamma^{i}(\bar{\omega},\bar{\omega}') + \beta^{-1}(\bar{\omega})\gamma^{i}(\bar{\omega},\omega_{\text{d}})\right)-\frac{1}{2} \Gamma_{\text{d}},\nonumber\\
\Gamma(\bar{\omega},\omega_{\text{d}})\!\!\!\!&=&\!\!\!\!-\frac{1}{2}\sum_{i=\text{r,nr}}\left( \gamma^{i}(\omega_{\text{d}},\omega_{\text{d}})+ \int_\chi^\infty d\bar{\omega}' \beta^{-1}(\bar{\omega}')  \gamma^{i}(\bar{\omega},\bar{\omega}')+ \beta^{-1}(\bar{\omega})\gamma^{i}(\bar{\omega},\omega_{\text{d}})\right)-\frac{1}{2} \Gamma_{\text{d}} ,\nonumber\\
\Gamma(\bar{\omega},\bar{\omega}')\!\!\!\!&=&\!\!\!\!-\frac{1}{2}\sum_{i=\text{r,nr}}\left[ \beta^{-1}(\bar{\omega})\gamma^{i}(\bar{\omega},\omega_{\text{d}})+\beta^{-1}(\bar{\omega}')\gamma^{i}(\bar{\omega}',\omega_{\text{d}}) + \int d\bar{\omega}'' \left(\beta^{-1}(\bar{\omega}) \gamma^{i}(\bar{\omega},\bar{\omega}'')+\beta^{-1}(\bar{\omega}') \gamma^{i}(\bar{\omega}',\bar{\omega}'')\right)\right].\nonumber
\end{eqnarray} 
The rate equation dramatically simplifies once we take the continuum limit $L\rightarrow \infty$, leading to all the terms in $\beta^{-1}(\bar{\omega})$ vanishing and leaving behind only the rate equation reported in Eq.~\ref{rate equation} in the main text.
 \begin{eqnarray} \label{re}
    \dot{N}_{\text{d}}&=&-\Gamma_{\text{d}} N_{\text{d}},\\
    \dot{N}_{\bar{\omega}}&=&\sum_{i=\text{r,nr}}\gamma^{i}(\omega_{\text{d}},\bar{\omega}) N_{\text{d}},\nonumber\\
    \dot{\rho}_{{\text{d}},\bar{\omega}}&=& -i (\omega_{\text{d}}-\bar{\omega})\rho_{{\text{d}},\bar{\omega}}+\Gamma_{\rho}\rho_{{\text{d}},\bar{\omega}} ,\nonumber\\
    \dot{\rho}_{\bar{\omega},{\text{d}}}&=& -i(\bar{\omega}-\omega_{\text{d}})\rho_{\bar{\omega},{\text{d}}}+\Gamma_{\rho}\rho_{\bar{\omega},{\text{d}}} ,\nonumber\\
\dot{\rho}_{\bar{\omega},\bar{\omega}'}|_{\bar{\omega}\neq \bar{\omega}'}&=&-i(\bar{\omega}-\bar{\omega}')\rho_{\bar{\omega},\bar{\omega}'}+\Gamma(\bar{\omega},\bar{\omega}') \rho_{\bar{\omega},\bar{\omega}'},\nonumber
\end{eqnarray}
with effective decay rates in the continuum limit
\begin{eqnarray}\label{rates}
\Gamma_{\text{d}}&=&  \sum_{i=\text{r,nr}} \int d\bar{\omega}\, \gamma^{i}(\omega_{\text{d}},\bar{\omega}),\\ \label{gammaC}
 \Gamma_{\rho}&=& -\frac{1}{2} \sum_{i=\text{r,nr}} \gamma^{i}(\omega_{\text{d}},\omega_{\text{d}})-\frac{1}{2} \Gamma_{\text{d}} ,\nonumber\\
\Gamma(\bar{\omega},\bar{\omega}')&=&0.\nonumber
\end{eqnarray} 
In the continuum limit the dynamics becomes irreversible, as from Fermi golden rule. The electron, once lost into the ionization continuum, cannot thus fall back into the bound state, and as such it doesn't contribute to the fluorescence spectrum of the system in Eq.~\ref{S}.
Defining $\Sigma^{i}(\omega_\lambda)$ the density of the reservoir for the channel $i$,
the emission rates can be explicitly calculated 
\begin{eqnarray} 
\gamma^{\text{r}}(\omega_{\text{d}},\omega_{\text{d}})\!\!\!\!&=&\!\!\!\!  2 \int_0^\infty d\omega_\lambda \Sigma^{\text{r}}(\omega_\lambda) \mu^2 \omega_\lambda d_{\text{b}}^2 {\Lambda^{\text{r}}_d}^2 \delta(\omega_\lambda-\omega_0) = 2  \Sigma^{\text{r}}(\omega_0) \omega_0 \mu^2 d_{\text{b}}^2 {\Lambda^{\text{r}}_{\text{d}}}^2, \\ 
\gamma^{\text{nr}}(\omega_{\text{d}},\omega_{\text{d}})\!\!\!\!&=&\!\!\!\!  2 \int_0^\infty d\omega_\lambda \Sigma^{\text{nr}}(\omega_\lambda) \nu^2 d_{\text{b}}^2 {\Lambda^{\text{nr}}_d}^2 \delta(\omega_\lambda-\omega_0) = 2  \Sigma^{\text{nr}}(\omega_0) \nu^2 d_{\text{b}}^2 {\Lambda^{\text{nr}}_{\text{d}}}^2, \\ 
  \gamma^{\text{r}}(\omega_{\text{d}},\bar{\omega})\!\!\!\!&=&\!\!\!\!  2 \int_0^\infty d\omega_\lambda \Sigma^{\text{r}}(\omega_\lambda) \mu^2 \omega_\lambda \left[d_{\text{b}}^2 {{\Lambda}^{\text{r}}(\bar{\omega})}^2 \delta(\bar{\omega}-\omega_{\text{d}}+\omega_0+\omega_\lambda)+c_{\text{b}}(\bar{\omega})^2 {\Lambda^{\text{r}}}_{\text{d}}^2  \delta(\bar{\omega}-\omega_{\text{d}}-\omega_0+\omega_\lambda) \right]\nonumber \\ \label{g_wd_w}
 \!\!\!\!&=&\!\!\!\!  2 \Sigma^{\text{r}}(\omega_{\text{d}}+\omega_0-\bar{\omega})  (\omega_{\text{d}}+\omega_0-\bar{\omega}) \mu^2 c_{\text{b}}(\bar{\omega})^2 {\Lambda^\text{r}}_{\text{d}}^2 \Theta(\omega_{\text{d}}+\omega_0-\bar{\omega}), \label{gr_wd_w} \\
 \gamma^{\text{nr}}(\omega_{\text{d}},\bar{\omega})\!\!\!\!&=&\!\!\!\!  2 \int_0^\infty d\omega_\lambda \Sigma^{\text{nr}}(\omega_\lambda) \nu^2 \left[d_{\text{b}}^2 {{\Lambda}^{\text{nr}}(\bar{\omega})}^2 \delta(\bar{\omega}-\omega_{\text{d}}+\omega_0+\omega_\lambda)+c_{\text{b}}(\bar{\omega})^2 {\Lambda^{\text{nr}}}_{\text{d}}^2  \delta(\bar{\omega}-\omega_{\text{d}}-\omega_0+\omega_\lambda) \right]\nonumber\\ 
\!\!\!\!&=&\!\!\!\! \label{gnr_wd_w}  2 \Sigma^{\text{nr}}(\omega_{\text{d}}+\omega_0-\bar{\omega}) \nu^2 c_{\text{b}}(\bar{\omega})^2 {\Lambda^\text{nr}}_{\text{d}}^2  \Theta(\omega_{\text{d}}+\omega_0-\bar{\omega}),
  \end{eqnarray}
where $\Theta$ is the Heaviside step functions, enforcing the positivity of the emitted excitation energy $\omega_\lambda$ and as such enabling or blocking the direct decay process from discrete to continuum states. From Eq.~\ref{gr_wd_w} and Eq.~\ref{gnr_wd_w}, we see that when $\omega_0<\chi/2$ ,  that necessarily implies $\omega_{\text{d}}+\omega_0<\chi$, both the emission rates $\gamma^{i}(\omega_{\text{d}},\bar{\omega})$ and their integral $\Gamma_{\text{d}}$ vanish. 
In the calculations reported in the main text, we assumed the standard free space density of states $\Sigma^{\text{r}}(\omega_\lambda)$, for the radiative reservoir , and an effective flat density for the non-radiative one $\Sigma^{\text{nr}}(\omega_\lambda)\propto  Q\omega_\lambda^0$ with $Q$ constant.
By inserting $\Sigma^{\text{r}}(\omega_\lambda)=\frac{\omega_\lambda^2 V}{\pi^2 c^3} $ and  $\mu=\sqrt{\frac{e^2}{2 \epsilon_0 V}} $ in  Eq. \ref{g_wd_w}, with quantization volume $V$ and speed of light $c$, we obtain the radiative emission rate 
\begin{eqnarray}
    \gamma^{\text{r}}(\omega_{\text{d}},\bar{\omega})\!&=&\!2e^2 \frac{\left(\omega_\textrm{d}+\omega_0-\bar{\omega}\right)^3 }{2 \epsilon_0\pi^2 c^3} \!c_{\text{b}}(\bar{\omega})^2 \!{\Lambda^\text{r}}_{\text{d}}^2  \Theta(\omega_{\text{d}}+\omega_0-\bar{\omega})\nonumber.\\
\end{eqnarray}
Unlike its radiative counterpart, the non radiative emission function $\gamma^{\text{nr}}(\omega_{\text{d}},\bar{\omega})$ becomes dependent on the coupled continuum  energy only via the coefficient $c_\textrm{b}(\bar{\omega})$
\begin{eqnarray}
    \gamma^{\text{nr}}(\omega_{\text{d}},\bar{\omega})=2  \tilde{\nu} c_{\text{b}}(\bar{\omega})^2 {\Lambda^\text{nr}}_{\text{d}}^2  \Theta(\omega_{\text{d}}+\omega_0-\bar{\omega}),
\end{eqnarray}
with $\tilde{\nu}=\nu^2 Q$.
For the numerical calculations in Fig.~\ref{Fig:3} (g-i)  and Fig.~\ref{Fig:4}, we employed $\tilde{\nu}=10^{-3} \chi$.

\section{Calculating the resonance fluorescence spectrum} \label{appD}

  The resonance fluorescence spectrum is calculated as the real part of the Fourier transformation of the correlation function of the dipole polarization operator, according to the Wienier-Khinchin theorem
\begin{eqnarray} \label{SA}
    S(\omega_{\text{f}})\!&=&\!
 \text{Re}\! \left [ \int_0^\infty  \!\!\text{lim}_{t\rightarrow\infty} \!\langle \sigma^+(t+\tau)  \sigma^-(t) \rangle e^{-i(\omega_{\text{f}}-\omega_0)\tau }d\tau\right]\nonumber,\nonumber\\
  \end{eqnarray}
where  the collective dipole polarization operator can be written as $\sigma^+= \int_0^\infty d\omega D(\omega) \sigma_{\omega, \text{b}}$ \cite{cohen-tannoudji_atom-photon_2008}. Note that we limit ourselves to dipoles involving the bound state in the definition of the resonance fluorescence, physically implying that our detector has a finite size focused on the electronic bound state. In the continuum $L\rightarrow\infty$ limit it will thus not detect relaxation processes happening between arbitrarily far free electrons.

By applying the inverse relation of Eqs. \ref{A1} and \ref{A2} , the correlation function in Eq.~\ref{S} can be expressed in terms of the dressed jump operators $P_{\omega_1,\omega_2}$
\begin{eqnarray}
 &&   \langle \sigma^+(\tau)  \sigma^- \rangle= \int_0^\infty d\omega D(\omega)  \int_0^\infty d{\omega'} D(\omega')  \langle \sigma_{\omega,\text{b}}(\tau)  \sigma_{\text{b},\omega'} \rangle=\\
    && d_{\text{b}}^2 {\Lambda^{\text{r}}_\text{d}}^2 \langle P_{\text{d},\text{d}}(\tau)  P_{\text{d},\text{d}}\rangle + \int d\bar{\omega}\int d\bar{\omega}'d_{\text{b}}^2 {\Lambda}^{\text{r}}(\bar{\omega}){\Lambda}^{\text{r}}(\bar{\omega}')e^{i (\bar{\omega}-\omega_{\text{d}}) \tau}\langle P_{\bar{\omega},\text{d}}(\tau)  P_{\text{d},\bar{\omega}'}\rangle+\nonumber\\
    && \int d\bar{\omega}\int d\bar{\omega}'c_{\text{b}}(\bar{\omega}) c_{\text{b}}(\bar{\omega}') {\Lambda^{\text{r}}_{\text{d}}}^2 e^{i (\omega_{\text{d}}-\bar{\omega}) \tau}\langle P_{\text{d},\bar{\omega}}(\tau)  P_{\bar{\omega}',\text{d}}\rangle + \nonumber\\
    &&\int d\bar{\omega}\int d\bar{\omega}'\int d\bar{\omega}'' c_{\text{b}}(\bar{\omega}') c_{\text{b}}(\bar{\omega}'') \bar{\Lambda}^{\text{r}}(\bar{\omega})^2 e^{i (\bar{\omega}-\bar{\omega}') \tau} \langle P_{\bar{\omega},\bar{\omega}'}(\tau)  P_{\bar{\omega}'',\bar{\omega}}\rangle \nonumber.
\end{eqnarray}

and evaluated by applying the Quantum Regression theorem \cite{ lax_quantum_1968,scully_quantum_1999,del_valle_regimes_2011}. By applying the rate equations in Eq.\ref{re1}-\ref{re5}, we can write

\begin{eqnarray} 
  \frac{d}{dt}  \langle P_{\text{d},\text{d}}(t)\rangle&=&\frac{d}{dt}\text{Tr}\left[P_{\text{d},\text{d}} \rho(t) \right]=\text{Tr}\left[P_{\text{d},\text{d}}\frac{dN_{\text{d}}}{dt}\right]=\\\nonumber
  &&
  \sum_{i=\text{r,nr}}\left[\int d\bar{\omega}\beta^{-1}(\bar{\omega}) {\gamma}^{i}(\bar{\omega},\omega_{\text{d}})\langle P_{\omega,\omega}(t) \rangle\right]- \Gamma_{\text{d}} \langle P_{\text{d},\text{d}}(t) \rangle,\\  
  \frac{d}{dt}  \langle P_{\text{d},\text{d}}(t)P_{\text{d},\text{d}}\rangle&=&-\Gamma_{\text{d}} \langle P_{\text{d},\text{d}}(t) P_{\text{d},\text{d}}\rangle,\nonumber\\
  \langle P_{\text{d},\text{d}}(t) 
P_{\text{d},\text{d}}\rangle&=&N_{\text{d}} e^{-\Gamma_{\text{d}} t}. \nonumber\label{Pdd}
\end{eqnarray}
Similarly, the other correlation functions are calculated as 
\begin{eqnarray} \label{Djw}
     \langle P_{\text{d},\bar{\omega}}(t)P_{\bar{\omega}',\text{d}}\rangle &=& N_{\text{d}}\delta(\bar{\omega}-\bar{\omega}')  e^{\Gamma(\bar{\omega},\omega_{\text{d}})\tau}\\ \label{Pwd}
     \langle P_{\bar{\omega},\text{d}}(t)P_{\text{d},\bar{\omega}'}\rangle &=& N_{\bar{\omega} }\,\beta^{-1}(\bar{\omega}) \delta(\bar{\omega}-\bar{\omega}' )  e^{\Gamma(\omega_{\text{d}},\bar{\omega})\tau}\nonumber\\ 
     \langle P_{\bar{\omega},\bar{\omega}'}(t)P_{\bar{\omega}'',\bar{\omega}}\rangle &=& \delta(\bar{\omega}'-\bar{\omega}'') N_{\bar{\omega}} e^{\Gamma(\bar{\omega}',\bar{\omega})\tau}.\label{Pww}\nonumber
\end{eqnarray}
Finally, inserting all Eq. \ref{Pdd} into Eq. \ref{S}, we obtain 
\begin{eqnarray} 
    S(\omega_{\text{f}})\!&=& \!S_{\text{d-d}}(\omega_{\text{f}})+\!S_{\text{c-d}}(\omega_{\text{f}})\!+\!S_{\text{d-c}}(\omega_{\text{f}})\!+\!S_{\text{c-c}}(\omega_{\text{f}}),\nonumber\\
\end{eqnarray}
with spectral components
\begin{eqnarray}
&&S_{\text{d}}(\omega_{\text{f}})=d_{\text{b}}(\omega_{\text{d}})^2 {\Lambda^{\text{r}}_{\text{d}}}^2, N_{\text{d}} \frac{ \Gamma_{\text{d}} }{(\omega_{\text{f}}-\omega_0)^2+\Gamma_{\text{d}}^2},\\
   &&S_{\text{c-d}}(\omega_{\text{f}})= d_{\text{b}}^2  \int d\bar{\omega} \frac{ \beta^{-1}(\bar{\omega}){\Lambda}^{\text{r}}(\bar{\omega})^2  N_{\bar{\omega}}\Gamma(\bar{\omega},\omega_{\text{d}})}{(\omega_{\text{f}}-\omega_0+\omega_{\text{d}}-\omega)^2+\Gamma(\bar{\omega},\omega_{\text{d}})},\nonumber\\
   &&S_{\text{d-c}}(\omega_{\text{f}})= {\Lambda^{\text{r}}_{\text{d}}}^2  N_{\text{d}}\int d\bar{\omega}  \frac{c_{\text{b}}(\bar{\omega})^2\Gamma(\omega_{\text{d}},\bar{\omega})}{(\omega_{\text{f}}-\omega_0-\omega_{\text{d}}+\omega)^2+\Gamma(\omega_{\text{d}},\bar{\omega})^2},
    \nonumber\\
   &&S_{\text{c-c}}(\omega_{\text{f}})=\int d\bar{\omega}\int d\bar{\omega}' \frac{ N_{\bar{\omega}}{ \Lambda}^{\text{r}}(\bar{\omega})^2   c_{\text{b}}(\bar{\omega}')^2\Gamma( \bar{\omega}',\bar{\omega})}{(\omega_{\text{f}}-\omega_0-\bar{\omega}-\omega')^2+\Gamma( \bar{\omega}',\bar{\omega})^2}.\nonumber
\end{eqnarray}
Finally, it is crucial to apply the continuum limit $L\rightarrow \infty$
\begin{eqnarray}
&&\text{lim}_{L\rightarrow \infty}    \Gamma(\omega_{\text{d}}, \bar{\omega})=0,\\
&& \text{lim}_{L\rightarrow \infty}    \Gamma( \bar{\omega},\omega_{\text{d}})=\Gamma_{\rho},\nonumber\\
&&\text{lim}_{L\rightarrow \infty}  \Gamma(\bar{\omega}, \bar{\omega}')=0.\nonumber
\end{eqnarray}
In this limit, we arrive to spectral components 
\begin{eqnarray}
&&S_{\text{d-d}}(\omega_{\text{f}})=d_{\text{b}}^2 {\Lambda^{\text{r}}_d}^2 N_{\text{d}} \frac{ |\Gamma_{\text{d}}| }{(\omega_{\text{f}}-\omega_0)^2+\Gamma_{\text{d}}^2},\\
   &&S_{\text{c-d}}(\omega_{\text{f}})=0,\nonumber\\
   &&S_{\text{d-c}}(\omega_{\text{f}})\!=  \!\!\int \!d\bar{\omega}\! \frac{ N_{\text{d}} c_{\text{b}}(\bar{\omega})^2 {\Lambda^{\text{r}}_\text{d}}^2|\Gamma_{\rho}|}{(\omega_{\text{f}}-\omega_0-\omega_{\text{d}}+\omega)^2+\Gamma_{\rho}^2}, \nonumber\\
   &&S_{\text{c-c}}(\omega_{\text{f}})=0.\nonumber
\end{eqnarray}
We notice that a larger \(d_{\text{b}}\) coefficient, meaning a more localised \(\Psi_{\text{d}}\) state, results in a stronger discrete-to-discrete spectral component \(S_{\text{d-d}}(\omega_{\text{f}})\), whereas a larger \(c_{\text{b}}(\bar{\omega})\), indicating greater hybridisation with the continuum, leads to a stronger discrete-to-continuum spectral component \(S_{\text{d-c}}(\omega_{\text{f}})\).

As a proof of the consistency of our theory, we can verify that, multiplying each spectral component for the radiative coefficient and the density of state $\Sigma^{\text{r}}$, and integrating over the whole range of positive emission energy $\omega_{\text{f}}$,
we recover to the total photon flux matching the radiative emission terms in the rate equations for the populations
\begin{eqnarray}
Z_{\text{d}}&=& {\gamma}^{\text{r}}(\omega_{\text{d}},\omega_{\text{d}}) N_{\text{d}},\\
      Z_{\text{b-c}}&=&  N_{\text{d}}\int d\bar{\omega} {\gamma}^{\text{r}}(\omega_{\text{d}},\bar{\omega}).\nonumber
\end{eqnarray}

It is also instructive to consider what would be obtained if, instead of the resonance fluorescence spectrum, we calculated the absorption spectrum. The absorption spectrum is proportional to the imaginary part of the susceptibility, $A(\omega)\propto \mathrm{Im}\,\mathcal{\zeta}(\omega)$, which is given by the Fourier transform of the dipole commutator
\begin{equation}
    \mathrm{Im}\,\zeta(\omega)
    = \int_0^{\infty} d\tau\, e^{i(\omega-\omega_0)\tau}
    \langle[\sigma^-(\tau),\sigma^+(0)]\rangle_{\mathrm{ss}}.
\end{equation}
The commutator structure removes all elastic contributions, such as the
$\Psi_\textrm{d} \rightarrow \Psi_\textrm{d}$ transition. The resulting absorption spectrum would therefore exhibit only the
inelastic signatures of the light-stabilised resonance: in particular, a broad absorption lineshape associated with the discrete $\Psi_\textrm{d} \rightarrow \Psi_{\bar\omega}$ transition, without any Rayleigh (elastic) component at the laser frequency.
Consequently, the inelastic bound-to-continuum absorption envelope, centered at
$\omega_\textrm{d} + \omega_0 - \bar{\omega}$ and broadened by
$\Gamma(\omega_\textrm{d},\bar{\omega})$, would persist, clearly reflecting the creation of the discrete resonance below the continuum.

\end{document}